\documentclass[aps,twocolumn,amsmath,letterpaper]{revtex4}
\usepackage{amssymb}
\usepackage{graphicx}
\usepackage{array}
\usepackage{hhline}
\usepackage{longtable}

\renewcommand{\thetable}{\Roman{table}} \thetable

\allowdisplaybreaks[1]

\begin{document}

\title{Inverted Berezinskii-Kosterlitz-Thouless Singularity and
High-Temperature Algebraic Order in an Ising Model on a Scale-Free
Hierarchical-Lattice Small-World Network}
\author{Michael Hinczewski$^{1}$ and A. Nihat Berker$^{1-3}$}

\affiliation{$^1$Feza G\"ursey Research Institute, T\"UBITAK -
Bosphorus University, \c{C}engelk\"oy 34680, Istanbul, Turkey}

\affiliation{$^2$Department of Physics, Ko\c{c} University, Sar\i
yer 34450, Istanbul, Turkey}

\affiliation{$^3$Department of Physics, Massachusetts Institute of
Technology, Cambridge, Massachusetts 02139, U.S.A.}

\begin{abstract}

We have obtained exact results for the Ising model on a hierarchical
lattice incorporating three key features characterizing many
real-world networks---a scale-free degree distribution, a high
clustering coefficient, and the small-world effect.  By varying the
probability $p$ of long-range bonds, the entire spectrum from an
unclustered, non-small-world network to a highly-clustered,
small-world system is studied.  Using the self-similar structure of
the network, we obtain analytic expressions for the degree
distribution $P(k)$ and clustering coefficient $C$ for all $p$, as
well as the average path length $\ell$ for $p=0$ and $1$. The
ferromagnetic Ising model on this network is studied through an
exact renormalization-group transformation of the quenched bond
probability distribution, using up to 562,500 renormalized
probability bins to represent the distribution.  For $p < 0.494$, we
find power-law critical behavior of the magnetization and
susceptibility, with critical exponents continuously varying with
$p$, and exponential decay of correlations away from $T_c$.  For $p
\geq 0.494$, in fact where the network exhibits small-world
character, the critical behavior radically changes:  We find a
highly unusual phase transition, namely an inverted
Berezinskii-Kosterlitz-Thouless singularity, between a
low-temperature phase with non-zero magnetization and finite
correlation length and a high-temperature phase with zero
magnetization and infinite correlation length, with power-law decay
of correlations throughout the phase.  Approaching $T_c$ from below,
the magnetization and the susceptibility respectively exhibit the
singularities of $\exp(-C/\sqrt{T_c-T})$ and $\exp(D/\sqrt{T_c-T})$,
with $C$ and $D$ positive constants.  With long-range bond strengths
decaying with distance, we see a phase transition with power-law
critical singularities for all $p$, and evaluate an unusually narrow
critical region and important corrections to power-law behavior that
depend on the exponent characterizing the decay of long-range
interactions.

PACS numbers: 89.75.Hc, 64.60.Ak, 75.10.Nr, 05.45.Df
\end{abstract}
\maketitle
\def\s{\rule{0in}{0.28in}}

\section{Introduction}
\setlength{\LTcapwidth}{\columnwidth}

Complex networks provide an intriguing avenue for tackling one of
the long-standing questions in statistical physics: how the
collective behavior of interacting objects is influenced by the
topology of those interactions.  Inspired by the diversity of
network structures found in nature, researchers in recent years have
investigated a variety of statistical models on networks with
real-world characteristics~\cite{AlbertBarabasi, Newman, DoroRev}.
Three empirically common network types have been the focus of
attention: networks with large clustering coefficients, where all
neighbors of a node are likely to be neighbors of each other;
networks with ``small-world'' behavior in the average shortest-path
length, $\ell \sim \ln(N)$, where $N$ is the number of nodes; and
those with a power-law (scale-free) distribution of degrees.  Since
the pioneering network models of
Watts-Strogatz~\cite{WattsStrogatz}, which exemplified the first two
properties, and Barab{\'asi}-Albert~\cite{BarabasiAlbert}, which
showed how the third could arise from particular mechanisms of
network growth, significant advances have taken place in
understanding how these properties affect statistical systems. The
Ising model has been studied on small-world
networks~\cite{Gitterman, BarratWeigt, Pekalski, Hong, Jeong}, along
with the $XY$ model~\cite{Kim}, and on Barab\'asi-Albert scale-free
networks~\cite{Aleksiejuk,Bianconi}.  On random graphs with
arbitrary degree distributions, the Ising model shows a range of
possible critical behaviors depending on the moments of the
distribution (or in the specific case of scale-free distributions,
the exponent describing the power-law tail)~\cite{Doro1,Leone}, a
fact which is accounted for by a phenomenological theory of critical
phenomena on these types of networks~\cite{Goltsev}.

In the current work we introduce a novel network structure based on
a hierarchical lattice~\cite{BerkerOstlund, Kaufman,Kaufman2}
augmented by long-range bonds.  By changing the probability $p$ of
the long-range bonds, we observe an entire spectrum of network
properties, from an unclustered network for $p=0$ with $\ell \sim
N^{1/2}$, to a highly-clustered small-world network for $p=1$ with
$\ell \sim \ln N$.  In addition, the network has a scale-free degree
distribution for all $p$.  Due to the hierarchical construction of
the network, together with the stochastic element introduced through
the attachment of the long-range bonds, this network combines
features of deterministic and random scale-free growing
networks~\cite{Barabasi2,Doro4,Doro3,ComellasSampels,Jung,ZhangRong,Andrade,Zhou,AndradeHerrmann},
and in the $p=1$ limit its geometrical properties are similar to the
pseudofractal graph studied in Ref.~\cite{Doro3}.  The self-similar
structure of the network allows us to calculate analytic expressions
for the degree distribution and clustering coefficient for all $p$,
as well as the average shortest-path length $\ell$ in the limiting
cases $p=0$ and $1$.

A renormalization-group transformation is formulated for the Ising
model on the network, yielding a variety critical behaviors of
thermodynamic densities and response functions. For the quenched
disordered system at intermediate $p$, we study the Ising model
through an exact renormalization-group transformation of the
quenched bond probability distribution, implemented numerically
using up to 562,500 renormalized probability bins to represent the
distribution. We find a finite critical temperature at all $p$, with
two distinct regimes for the critical behavior.  When $p < 0.494$,
the magnetization and susceptibility show power-law scaling, and
away from $T_c$ correlations decay exponentially, as in a typical
second-order phase transition.  The magnitudes of the critical
exponents, which continuously vary with $p$, become infinite as $p
\to 0.494$ from below.  For $p \geq 0.494$, in fact coinciding with
the onset of the small-world behavior of the underlying network, we
find a highly unusual infinite-order phase transition: an inverted
Berezinskii-Kosterlitz-Thouless
singularity~\cite{Berezinskii,KosterlitzThouless}, between a
low-temperature phase with non-zero magnetization and finite
correlation length, and a high-temperature phase with zero
magnetization and infinite correlation length, exhibiting power-law
decay of correlations (in contrast to the typical
Berezinskii-Kosterlitz-Thouless phase transition, where the
algebraic order is in the low-temperature phase).  Approaching $T_c$
from below, the magnetization and the susceptibility respectively
behave as $\exp(-C/\sqrt{T_c-T})$ and $\exp(D/\sqrt{T_c-T})$, with
$C$ and $D$ calculated positive constants.

Infinite-order phase transitions have been observed for the Ising
model on random graphs with degree distributions $P(k)$ that have a
diverging second moment $\langle k^2 \rangle$~\cite{Doro1,Leone},
but for these systems $T_c = \infty$ on an infinite network.  An
infinite-order percolation transition has been seen in models of
growing
networks~\cite{Callaway,Doro5,Kim2,Lancaster,Bauer2,Coulomb,Bollobas,Krapivsky},
with exponential scaling in the size of the giant component above
the percolation threshold.  A prior observation of a
finite-temperature, inverted Berezinskii-Kosterlitz-Thouless
singularity similar to the one described above was made in
Ising models on an inhomogeneous growing network~\cite{Bauer} and on
a one-dimensional inhomogeneous lattice~\cite{Costin1,Costin2,Bundaru,Romano}.

The final aspect of our network we investigated was the effect of
adding distance-dependence to the interaction strengths of the
long-range bonds, along the lines of Ref.~\cite{Jeong}, where
distance-dependent interactions were considered in a small-world
Ising system.  With decaying interactions, the second-order phase
transition for all $p$ has a strongly curtailed critical region and
corrections to power-law behavior that vary with the exponent
$\sigma$ describing the decay of interactions.

This paper is organized as follows:  In Sec.~II, we describe
the basic geometric features of the network, starting with the
construction of the lattice (II.A), whose deterministic nature
allows us to derive exact results on the degree distribution
(II.B.1), clustering coefficient (II.B.2), and average shortest-path
length (II.B.3).  Additional details about the derivations in this
section are included in Appendix A.  In Sec.~III, we study the Ising
model on the network through exact renormalization-group
transformations, first for the simpler cases where the long-distance
bonds are either all absent (III.A), or all present (III.B).  In the
latter case, we look at both uniform interaction strengths along the
long-distance bonds (III.B.1) and interactions decaying with
distance (III.B.2).  Finally, we turn to the more complex situation
where there is a quenched random distribution of the long-distance
bonds in the lattice, which requires formulating a
renormalization-group transformation in terms of quenched
probability distributions (III.C.1).  Analysis of the flows of these
distributions gives a complete picture of the thermodynamics of our
system over the entire range of $p$ (III.C.2).  We present our
conclusions in Sec.~IV.

\section{Hierarchical-Lattice Small-World Network}

\subsection{Construction of the Lattice}

\begin{figure}[h]
\includegraphics*[scale=0.49]{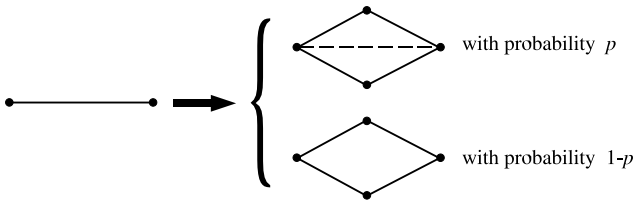}
\caption{Construction of the hierarchical lattice.  The solid lines
  correspond to nearest-neighbor bonds, while the dashed lines are
  long-range bonds, which occur with probability p.}\label{fig1}
\end{figure}

\begin{figure}[h]
\centering\includegraphics*[scale=0.42]{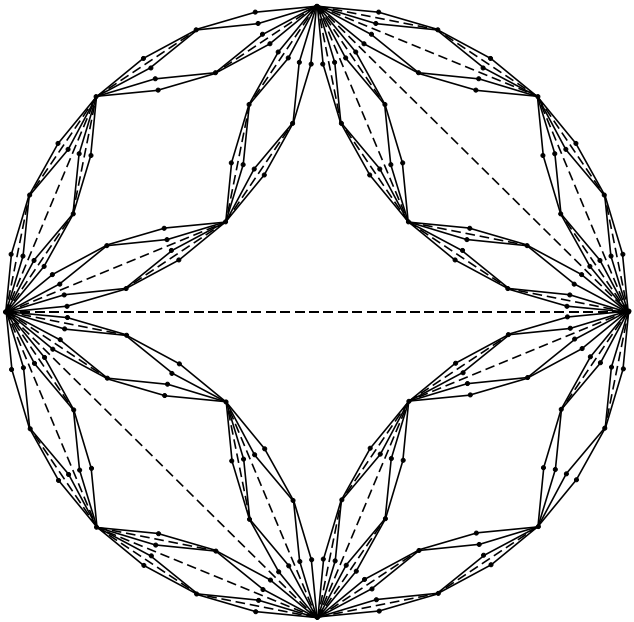}
\caption{An example of the hierarchical lattice after $n=4$ steps in
  the construction, for $p=0.6$.}\label{fig2}
\end{figure}

We construct a hierarchical lattice~\cite{BerkerOstlund, Kaufman,
Kaufman2} as shown in Fig.~\ref{fig1}.  The lattice has two types of
bonds: nearest-neighbor bonds (depicted as solid lines) and
long-range bonds (depicted as dashed lines).  In each step of the
construction, every nearest-neighbor bond is replaced either by the
connected cluster of bonds on the top right of Fig.~\ref{fig1} with
probability $p$, or by the connected cluster on the bottom right
with probability $1-p$.  This procedure is repeated $n$ times, with
the infinite lattice obtained in the limit $n \to \infty$.  The
initial ($n=0$) lattice is two sites connected by a single
nearest-neighbor bond.  An example of the lattice at $n=4$ for an
arbitrary $p \ne 0,1$ is shown in Fig.~\ref{fig2}.

The $p=0$ case, with no long-range bond, is the hierarchical lattice
~\cite{BerkerOstlund} on which the
Migdal-Kadanoff~\cite{Migdal,Kadanoff} recursion relations with
dimension $d=2$ and length rescaling factor $b=2$ are exact.  As
will be seen below, the network in this case exhibits no small-world
feature, with a clustering coefficient $C=0$ and an average
shortest-path length $\ell$ that scales like $N^{1/2}$, where $N$ is
the number of sites in the lattice.  The $p=1$ case, on the other
hand, shows typical small-world properties, with the presence of
long-range bonds giving the high clustering coefficient $C = 0.820$
and an average path length which scales more slowly with system
size, $\ell \sim \ln N$.  By varying the parameter $p$ from 0 to 1,
we continuously move between the two limits.  These and other
network characteristics of our hierarchical lattice are discussed in
detail in the next section.

\subsection{Network Characteristics}

\subsubsection{Degree Distribution}

After the $n$th step of the construction, there are a total of $N_n
= \frac{2}{3}(2+4^n)$ sites in the lattice.  We categorize these
sites by the number of nearest-neighbor bonds attached to the site,
$k_\text{nn}$, and the maximum possible number of long-range bonds
attached to the site, $k_\text{ld}$, of which on average only
$pk_\text{ld}$ actually exist.  At the $m$th level there are
$4^{n-m+1/2}$ sites with $k_\text{nn}=2^m$, $k_\text{ld}=2^m-2$, for
$m=1,\ldots,n$.  In addition, there are two sites with
$k_\text{nn}=2^n$, $k_\text{ld}=2^n-1$.  Thus, the non-zero
probabilities that a randomly chosen site has degree $k$ are
\begin{equation}\label{eq:2}
P_n(k) = \begin{cases} \frac{4^{n-m+1/2}}{N_n} \binom{2^m-2}{r}p^r (1-p)^{2^m-2-r},\\
\frac{2}{N_n} \binom{2^n-2}{r} p^r (1-p)^{2^n-2-r}\\
\qquad+\frac{2}{N_n} \binom{2^n-1}{r} p^r (1-p)^{2^n-1-r},\\
\frac{2}{N_n} p^{2^n-1},
\end{cases}
\end{equation}
respectively for
\begin{align}\label{eq:2a}
k=&2^m+r,\: 0\le r \le 2^m-2,\: 1\le m \le n-1,\nonumber\\
k=&2^n+r,\:0\le r \le 2^n-2,\nonumber\\
k=&2^{n+1}-1.
\end{align}

Since the degree distribution is not continuous, the exponent
describing the power-law decay of degrees is extracted from the
cumulative distribution function~\cite{Newman} in the $n \to \infty$
limit, $P_\text{cum}(k) = \sum_{k^\prime = k}^{\infty} P(k^\prime)$,
where $P(k) = \lim_{n \to \infty} P_n(k)$.  For a scale-free network
of exponent $\alpha$, $P_\text{cum}(k) \sim k^{1-\alpha}$. In our
case $P_\text{cum}(k) \sim k^{-2}$ for large $k$, giving $\alpha =
3$, a value comparable to the exponents of many real-world
scale-free networks~\cite{AlbertBarabasi}.  The maximum degree
$k_\text{max}$ in the scale-free network should scale as
$k_\text{max} \sim N^{1/(\alpha-1)}_n$~\cite{Newman}, which is
indeed satisfied, for large $n$, in our network.  The average degree
$\langle k \rangle_n$ after $n$ construction steps is
\begin{equation}\label{eq:2b}
\langle k \rangle_n = \sum_{k=1}^{\infty} k P_n(k) = 3+p -\frac{3(2+p)}{2+4^n}\,,
\end{equation}
which goes to $\langle k \rangle = 3+p$ in the infinite lattice limit.

\subsubsection{Clustering Coefficient $C_m$}

\begin{figure}[t]
\centering\includegraphics*[scale=1]{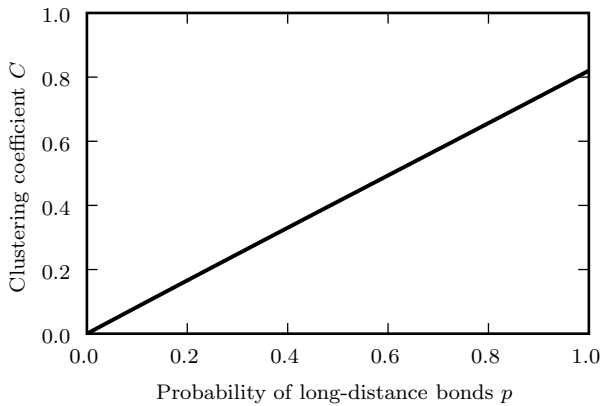}
\caption{Clustering coefficient $C$ for the infinite lattice as a
function of the probability of long-range bonds $p$.}\label{fig3}
\end{figure}

If a given site in the network is connected to $k$ sites, defined as
the neighbors of the given site, the ratio between the number of
bonds among the neighbors and the maximum possible number of such
bonds $k(k-1)/2$ is the clustering coefficient of the given
site~\cite{WattsStrogatz}. The clustering coefficient $C$ of the
network is the average of this coefficient over all the sites, and
can take on values between 0 and 1, the latter corresponding to a
maximally clustered network where all neighbors of a site are also
neighbors of each other.  For our network in the $n \to \infty$
limit, $C$ can be evaluated exactly:  The fraction $\lim_{n \to
\infty} 4^{n-m+1/2}/N_n = 3\cdot 4^{-m}$ of the sites, with
$k_\text{nn} = 2^m$ and $k_\text{ld} = 2^m-2$, have the average
clustering coefficient $C_m$, where $C_1 = p$ and $C_m$ for $m > 1$
is, as derived in Appendix A.1,
\begin{multline}\label{eq:3}
C_m = \sum_{r=0}^{2^{m-1}}\sum_{r^\prime=0}^{2^{m-1}-2}
\binom{2^{m-1}}{r} \binom{2^{m-1}-2}{r^\prime} \cdot
\\ \frac{2 p^{r+r^\prime}(1-p)^{2^m-2-r-r^\prime}\left\{ 2r+p\binom{r+r^\prime}{2}\frac{2^m-3}{\binom{2^m-2}{2}}\right\}}{(2^m+r+r^\prime)(2^m+r+r^\prime-1)}\,.
\end{multline}
We plot the clustering coefficient $C$
\begin{equation}\label{eq:4}
C = \sum_{m=1}^{\infty} 3\cdot 4^{-m} C_m\,.
\end{equation}
as a function of $p$ in Fig.~\ref{fig3}.  Note that $C$ increases
almost linearly from $0$ at $p=0$ to $0.820$ at $p=1$, as can also
be seen from the expansion of Eq.~\eqref{eq:4} to second order in
$p$,
\begin{equation}\label{eq:5}
C = 0.837p-0.0378p^2+\text{O}(p^3)\,.
\end{equation}

\subsubsection{Average Shortest-Path Length $\ell_n$}

Let $d_{ij}$ be the shortest-path length between two sites $i$ and
$j$ in the network, measured in terms of the number of bonds along
the path.  The average shortest-path length $\ell_n$ is the average
of $d_{ij}$ over all pairs of sites $i$,$j$ at the $n$th level. For
general $p$ we have evaluated this quantity numerically. For $p=0$
and $p=1$ we have obtained exact analytic expressions (Appendix
A.2), revealing qualitatively distinct behaviors:  For $p=0$ we find
\begin{align}\label{eq:6}
\ell_n &= \frac{2^n (98+27\cdot 2^n +42\cdot 4^n + 22\cdot 16^n+21n\cdot 4^n)}{21(2+5\cdot 4^n +2\cdot 16^n)}\nonumber\\
&\xrightarrow[n\to \infty]{} \frac{11}{21}2^n\,,
\end{align}
and since $N_n \sim 4^n$ for large $n$, we have $\ell_n \sim
N_n^{1/2}$.  Comparing this result to that of a hypercubic lattice
of dimension $d$, where the average shortest-path length scales as
$N^{1/d}$~\cite{AlbertBarabasi}, we see that $\ell_n$ for the $p=0$
network has the power-law scaling behavior of the square lattice.
For $p=1$, on the other hand, we find
\begin{align}\label{eq:7}
&\ell_n =\nonumber\\
&\frac{23+4\cdot(-2)^n +44\cdot 4^n +10\cdot 16^n +6n\cdot
4^n+12n\cdot 16^n}
{9(2+5\cdot 4^n+2\cdot16^n)}\nonumber\\
&\xrightarrow[n\to \infty]{} 2n/3~,
\end{align}
which means that $\ell_n \sim \ln(N_n)$ for large $n$.  This much
slower, logarithmic scaling of $\ell_n$ with lattice size, together
with the high clustering coefficient, are the defining features of a
small-world network.

\begin{figure}[t]
\centering\includegraphics*[scale=1]{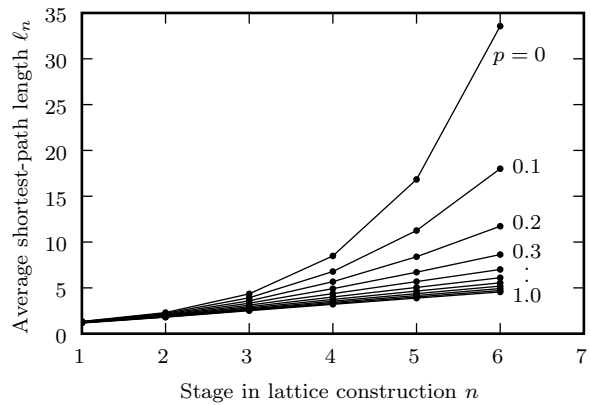}
\caption{Average shortest-path length $\ell_n$ for level $n$, shown
for various values of $p$ between 0 and 1. For $p=0$ and $p=1$,
$\ell_n$ is given exactly by Eqs.~\eqref{eq:6} and~\eqref{eq:7}. For
other $p$, we have calculated $\ell_n$ numerically, with an accuracy
of $\pm 0.3\%$.}\label{fig4}
\end{figure}

In Fig.~\ref{fig4} we show $\ell_n$ calculated for for the full
range of $p$ between 0 and 1, for $n$ up to 6.  It is evident that
even a small percentage of long-range bonds drastically reduces the
average shortest-path length, and that $\ell_n$ shows small-world
characteristics, scaling nearly linearly with $n$, for $p \gtrsim
0.5$.  We shall see below that the small-world structure at larger
$p$ translates into a distinctive critical behavior for the Ising
model on this network.

\section{Ising Model on the Network}

We study the Ising model on the network introduced in the previous
section, with Hamiltonian
\begin{align}
-\beta {\cal H} =& J \sum_{\langle i j \rangle_\text{nn}} s_i s_j + \sum_{\langle i j \rangle_\text{ld}} K_{ij} s_i s_j \nonumber\\
&+ H_B \sum_{\langle i j \rangle_\text{nn}} (s_i + s_j) + H_S \sum_i s_i \label{eq:8}\,,
\end{align}
where $J, K_{ij}>0$, $\langle i j \rangle_\text{nn}$ denotes
summation over nearest-neighbor bonds, and $\langle i j
\rangle_\text{ld}$ denotes summation over long-range bonds. We
generalize the above, by introducing a distance dependence in the
interaction constants $K_{ij}$,
\begin{equation}\label{eq:9}
K_{ij} = J m_{ij}^{-\sigma}\,.
\end{equation}
Here the exponent $\sigma \ge 0$, and $m_{ij}$ measures the range of
the long-range bond between sites $i$ and $j$:  For a lattice
constructed in $n$ steps, those long-range bonds that appear at the
$n$th step have $m_{ij} = 1$, those that appear at the $(n-1)$th
step have $m_{ij} = 2$, and so on until the long-range bond that
appears at the first step, which has $m_{ij} = n$. The long-range
term in the Hamiltonian can be rewritten as
\begin{equation}\label{eq:9aa}
\sum_{\langle i j \rangle_\text{ld}} K_{ij} s_i s_j = K_1
\sum_{\langle i j \rangle_{\text{ld},1}}  s_i s_j + K_2
\sum_{\langle i j \rangle_{\text{ld},2}}  s_i s_j + \cdots\,,
\end{equation}
where $K_{q} \equiv J q^{-\sigma}$ and $\langle i j
\rangle_{\text{ld},q}$ denotes summation over long-range bonds with
$m_{ij} = q$.

The Hamiltonian of Eq.~\eqref{eq:8} includes two types of magnetic
field terms, one counted with bonds ($H_B$) and the other counted
with sites ($H_S$).  We shall calculate the associated spontaneous
magnetizations at $H_B = H_S = 0$,
\begin{equation}\label{eq:9b}
M_B = \frac{1}{N_\text{nn}} \sum_{\langle i j \rangle_\text{nn}}
\langle s_i + s_j \rangle\,,\quad M_S = \frac{1}{N_\text{n}} \sum_i
\langle s_i \rangle\,,
\end{equation}
where $N_\text{nn}= 4^n$ is the number of nearest-neighbor bonds
after the $n$th construction stage, so that $N_\text{nn}/N_\text{n}
= 3/2$ in the limit $n \to \infty$.  For a translationally invariant
lattice, where each site has the same degree, $M_B$ and $M_S$ would
be simply related by $M_B = 2 M_S$, but for the hierarchical lattice
this is no longer true due to the different degrees of the sites.

Before turning to the phase diagram and critical properties of the
system for general $p$, which require formulating a
renormalization-group transformation in terms of quenched
probability distributions, we present the distinct critical
behaviors of the limiting cases of $p=0$ and $p=1$.

\begin{table*}[t]
\renewcommand{\arraystretch}{1.3}
\begin{tabular}{|cc|c|c|c|c|c|c|}
\hline
\multicolumn{2}{|c|}{Property} & $p=0$ &\parbox{0.85in}{\vspace{0.3em}$0 < p <
  0.494$, $\sigma = 0$\vspace{0.3em}} &
\parbox{0.85in}{\vspace{0.3em}$0.494 \le p \le 1$, $\sigma =
  0$\vspace{0.3em}} & \parbox{0.85in}{$0< p < 1$, $0<\sigma < 1$} & \parbox{0.85in}{$p=1$, $0<\sigma < 1$} &  \parbox{0.85in}{$0<p\le 1$, $\sigma \ge 1$}\\
\hline\hline
\multicolumn{2}{|c|}{$T_c$} & 1.641 & \parbox{0.75in}{varies with $p$
  (see Fig.~\ref{fig:temps}); reaches 3.592 at $p=0.494$}& \parbox{0.75in}{varies with $p$
  (see Fig.~\ref{fig:temps}); reaches 7.645 at $p=1$ } &
\parbox{0.75in}{varies with $\sigma$\\ and $p$\\ (see
  Fig.~\ref{fig:p1tc})} & \parbox{0.75in}{varies with $\sigma$
  (see Fig.~\ref{fig:p1tc});\\ reaches 3.485 at $\sigma=1$ } & \parbox{0.75in}{varies with $\sigma$\\
  and $p$\\ (see Fig.~\ref{fig:p1tc})} \\\hline
\multicolumn{2}{|c|}{$y_T$} & 0.747 & vary with $p$ & 0 & 0.747 & 0.747 &0.747 \\
\multicolumn{2}{|c|}{$y_H$} & 1.879 & (see Fig.~\ref{fig:der}) & 1.585
&1.879 &1.879 &1.879\\\hline
\multicolumn{2}{|c|}{$\xi$}  & $|t|^{-1/y_T}$ & $|t|^{-1/y_T}$ &
$e^{A/\sqrt{|t|}} \, (t<0)$ & $|t|^{-\frac{1}{y_T}+f_1(p,\sigma,t)}$ & $|t|^{-\frac{1}{y_T}-C_1\left(-\ln |t|\right)^{-\sigma}}$ &$|t|^{-1/y_T}$\\
    &  &  &  & $\infty \, (t>0)$ & &  & \\\hline
\multicolumn{2}{|c|}{$C_{\text{\scriptsize{ sing}}}$}  & $|t|^{-\frac{2y_T
  -2}{y_T}}$ & $|t|^{-{2y_T
  -2}{y_T}}$ &
$|t|^{-3/2} e^{-2A/\sqrt{|t|}}$ &
$|t|^{-\frac{2y_{T}-2}{y_{T}}+f_2(p,\sigma,t)}$&
$|t|^{-\frac{2y_{T}-2}{y_{T}}+2C_1\left(-\ln |t|\right)^{-\sigma}}$
& $|t|^{-\frac{2y_T -2}{y_T}}$\\
\hline $M_B$, $M_S$ & $(t<0)$ &
$|t|^{\frac{2-y_H}{y_T}}$ & $|t|^{\frac{2-y_H}{y_T}}$ & $e^{-A(2-y_H)/\sqrt{|t|}}$ &
$|t|^{\frac{2-y_H}{y_T}+f_3(p,\sigma,t) }$ &
$|t|^{\frac{2-y_H}{y_T}+C_2\left(-\ln |t|\right)^{-\sigma} }$
&$|t|^{\frac{2-y_H}{y_T}}$\\
\hline \parbox{0.7in}{$\chi_{BB}$, $\chi_{BS}$, $\chi_{SS}$} &
$(t<0)$ & $|t|^{-\frac{2y_H-2}{y_T}}$ & $|t|^{-\frac{2y_H-2}{y_T}}$ & $e^{A(2y_H-2)/\sqrt{|t|}}$ &
 $|t|^{-\frac{2y_H-2}{y_T} +f_4(p,\sigma,t)}$ &
 $|t|^{-\frac{2y_H-2}{y_T} +C_3\left(-\ln |t|\right)^{-\sigma}}$ & $|t|^{-\frac{2y_H-2}{y_T}}$\\

\hline
\end{tabular}\caption{Critical properties of the hierarchical-lattice network for all cases of long-range bond probabilities $p$ and long-range
  bond decay exponents $\sigma$.  The functions $f_i(p,\sigma,t)$ are
  corrections to the $p=0$ scaling behavior, which are significant at
  larger $p$.
  The factor $A$ is defined below Eq.~\eqref{eq:29}, and
  $C_1$, $C_2$, $C_3$ are given as functions of $y_T,y_H,\sigma$ in
  Eqs.~\eqref{eq:60}.}

\label{tab1}\end{table*}

\subsection{Critical Properties at $p=0$}

The $d=2$, $b=2$ Migdal-Kadanoff recursion relations are exact
~\cite{BerkerOstlund} on the $p=0$ lattice, and the
renormalization-group transformation consists of decimating the two
center sites in the cluster shown on the bottom right of
Fig.~\ref{fig1}. The renormalized Hamiltonian of the two remaining
sites $i^\prime$, $j^\prime$ is
\begin{equation}
-\beta {\cal H^\prime} =
 \sum_{\langle i^\prime j^\prime \rangle} \left[J^\prime  s_{i^\prime} s_{j^\prime}
 + H_B^\prime (s_{i^\prime} + s_{j^\prime}) + G^\prime\right]
 + H_S^\prime \sum_{i^\prime} s_{i^\prime}\label{eq:88}\,,
\end{equation}
where the renormalized interaction constants are~\cite{McKayBerker}:
\begin{align}
J^\prime &= \frac{1}{2}\ln \left(R_{++}R_{--}/R_{+-}^2\right)\,,\nonumber\\
H_B^\prime &= \frac{1}{2}\ln \left(R_{++}/R_{--}\right)\,,\qquad H_S^\prime = H_S\,,\nonumber\\
G^\prime &= 4 G + \frac{1}{2}\ln
\left(R_{++}R_{--}R_{+-}^2\right)\,,\label{eq:10}
\end{align}
with
\begin{align}
R_{++} &= x y^2 z + x^{-1} z^{-1}\,,\quad R_{--} = x^{-1} z + x y^{-2} z^{-1}\,,\nonumber\\
R_{+-} &= y z + y^{-1} z^{-1}\,,\quad x = e^{2J}\,,\: y =
e^{2H_B}\,,\: z=e^{H_S}\,.\label{eq:10b}
\end{align}
Here $G$ is an additive constant per bond, equal to zero in the
original Hamiltonian, but always generated by the transformation and
necessary for the calculation of densities and response functions.
From the transformation in Eqs.~\eqref{eq:10},\eqref{eq:10b} we see
that an initial Hamiltonian with only an $H_S$ magnetic field term
will invariably generate an $H_B$ term upon renormalization.

The subspace $H_B = H_S =0$ is up-down symmetric in spin space and
closed under the transformation. Within this subspace, there is one
unstable fixed point at
\begin{equation}\label{eq:11}
J_c =
\ln\left[\frac{1}{3}\left(1+(19-3\sqrt{33})^{1/3}+(19+3\sqrt{33})^{1/3}\right)\right]\,,
\end{equation}
corresponding to a temperature $T_c = 1/J_c = 1.641$.  Under
renormalization-group transformations, the system renormalizes at
high temperatures $J < J_c$ to the sink at $J^\ast = 0$ of the
disordered phase and at low temperatures $J > J_c$ to the sink at
$J^\ast = \infty$ of the ordered phase.  The critical behavior at
$T_c$ is obtained from the eigenvalues of the recursion matrix at
the critical fixed point,
\begin{multline}\label{eq:12}
\begin{pmatrix}  \frac{\partial J^\prime}{\partial J} &\frac{\partial J^\prime}{\partial H_B}
&\frac{\partial J^\prime}{\partial H_S}\\
\frac{\partial H_B^\prime}{\partial J} &\frac{\partial
H_B^\prime}{\partial H_B} &\frac{\partial H_B^\prime}{\partial H_S}\\
\frac{\partial H_S^\prime}{\partial J} &\frac{\partial
H_S^\prime}{\partial H_B} &\frac{\partial H_S^\prime}{\partial H_S}
\end{pmatrix}
 =
\begin{pmatrix} 2u & 0 & 0 \\ 0 & 2+2u & u \\
0 & 0 & 1 \end{pmatrix}\,,
\end{multline}
where $u = \tanh 2J_c$.  This recursion matrix has eigenvalues $2u
\equiv b^{y_T}$, $2+2u \equiv b^{y_H}$, and 1, with eigenvalue
exponents $y_T = 0.747$, $y_H = 1.879$. Along the corresponding
eigendirections are one thermal and two magnetic scaling fields: $t
= \frac{J_c-J}{J_c} = \frac{T-T_c}{T_c}$, $h_1 = (2+\coth 2J_c)H_B +
H_S$, and $h_2 = H_S$, with linearized recursion relations $t^\prime
= b^{y_T} t$, $h_1^\prime = b^{y_H} h_1$, and $h_2^\prime = h_2$.
Standard eigenvalue analysis at the fixed point yields the critical
behaviors for the internal energy $U = \frac{1}{N_\text{nn}}
\sum_{\langle i j \rangle_\text{nn}} \langle s_i s_j \rangle$, the
magnetizations $M_B$, $M_S$, and the correlation length $\xi$:
\begin{align}
U - U_c &\sim |t|^{1-\alpha}\,,\qquad \alpha = \frac{2y_T -d}{y_T} = -0.677\,,\nonumber\\
M_S, M_B &\sim |t|^\beta\,\:\:\: (t<0)\,,\quad \beta = \frac{d-y_H}{y_T} = 0.162\,,\nonumber\\
\xi &\sim |t|^{-\nu}\,, \qquad \nu = \frac{1}{y_T} =
1.338\,.\label{eq:22}
\end{align}
$M_B$ and $M_S$ have the same critical exponent $\beta$, because the
dominant magnetic scaling field $h_1$ mixes $H_B$ and $H_S$.
Similarly, the susceptibility critical exponent is $\gamma = (2y_H
-d)/y_T = 2.353$.  Approaching criticality in the ordered phase, all
three susceptibilities one can define, $\chi_{BB} = \frac{\partial
M_B}{\partial H_B}$, $\chi_{BS} =
\sqrt{\frac{N_\text{nn}}{N_\text{n}}} \frac{\partial M_B}{\partial
H_S}$, and $\chi_{SS} = \frac{\partial M_S}{\partial H_S}$, have the
critical behavior $|t|^{-\gamma}$.  The zero-field susceptibilities
are infinite throughout the disordered phase.  To recall this, we
briefly review the calculation of thermodynamic densities and
response functions by multiplications along the
renormalization-group trajectory.

Let $\mathbf{K} = (G, J, H_B, H_S)$ be the vector of interaction
constants in the Hamiltonian, and $\mathbf{K}^\prime = (G^\prime,
J^\prime, H_B^\prime, H_S^\prime)$ the analoguous vector for the
renormalized system.  Corresponding to each component $K_\alpha$ of
$\mathbf{K}$ is a thermodynamic density $M_\alpha =
\frac{1}{N_\alpha} \frac{\partial \ln Z}{\partial K_\alpha}$, where
$Z$ is the partition function, and $N_\alpha$ is a component of the
vector $\mathbf{N} = (N_\text{nn}, N_\text{nn}, N_\text{nn}, N_n)$.
Thus, the density vector $\mathbf{M} = (1,U,M_B,M_S)$ is related to
the density vector of the renormalized system $\mathbf{M}^\prime$ by
the conjugate recursion relations~\cite{BerkerOstlundPutnam}:
\begin{equation}\label{eq:13}
M_\alpha = b^{-d} \sum_\beta M_\beta^\prime T_{\beta\alpha}\,,\qquad
T_{\beta\alpha} = \frac{N_\beta}{N_\alpha} \frac{\partial
K_\beta^\prime}{\partial K_\alpha}\,.
\end{equation}

An analogous recursion relation for response functions
$\chi_{\alpha\beta} = \sqrt{\frac{N_\alpha}{N_\beta}} \frac{\partial
M_\alpha}{\partial K_\beta}$ has been derived by McKay and
Berker~\cite{McKayBerker}:
\begin{align}
\chi_{\alpha\beta} =& b^{-d} \left[ \sum_{\lambda,\mu} \sqrt{\frac{N_\lambda N_\mu}{N_\alpha N_\beta}} \chi^\prime_{\lambda\mu} \frac{\partial K^\prime_\lambda}{\partial K_\alpha} \frac{\partial K^\prime_\mu}{\partial K_\beta}\right.\nonumber\\
&\qquad+ \left.\sum_\lambda \frac{N_\lambda}{\sqrt{N_\alpha N_\beta}} M^\prime_\lambda \frac{\partial^2 K^\prime_\lambda}{\partial K_\alpha \partial K_\beta}\right]\,.\label{eq:14}
\end{align}
Using the density-response vector $\mathbf{V} = (1,U,M_B,M_S,$ $
\chi_{BB},\chi_{BS},\chi_{SS})$, Eqs.~\eqref{eq:13} and
\eqref{eq:14} are combined into a single recursion relation,
\begin{equation}\label{eq:15}
V_\alpha = b^{-d} \sum_\beta V_\beta^\prime W_{\beta\alpha}\,.
\end{equation}
The extended recursion matrix $\overleftrightarrow{\mathbf{W}}$ for
the subspace $H_B = H_S = 0$ is
\begin{equation}
\small{\begin{pmatrix} b^{d} & \frac{\partial G^\prime}{\partial J} & 0 & 0 &\vline&  \frac{\partial^2 G^\prime}{\partial H_B^2} & \mu \frac{\partial^2 G^\prime}{\partial H_B \partial H_S} & \mu^2\frac{\partial^2 G^\prime}{\partial H_S^2} \\
 0 & \frac{\partial J^\prime}{\partial J} & 0 & 0 &\vline&  \frac{\partial^2 J^\prime}{\partial H_B^2} & \mu \frac{\partial^2 J^\prime}{\partial H_B \partial H_S} & \mu^2\frac{\partial^2 J^\prime}{\partial H_S^2} \\
 0 & 0 & \frac{\partial H_B^\prime}{\partial H_B} & \mu^2 \frac{\partial H_B^\prime}{\partial H_S} &\vline & 0 & 0 & 0 \\
 0 & 0 & 0 & \frac{\partial H_S^\prime}{\partial H_S} & \vline & 0 & 0 & 0\\
 \hline 0 & 0 & 0 & 0 & \vline & \left(\frac{\partial H_B^\prime}{\partial H_B}\right)^2 & \mu\frac{\partial H_B^\prime}{\partial H_B}\frac{\partial H_B^\prime}{\partial H_S} & \mu^2 \left(\frac{\partial H_B^\prime}{\partial H_S}\right)^2\\
 0 & 0 & 0 & 0 & \vline & 0 & \frac{\partial H_B^\prime}{\partial H_B}\frac{\partial H_S^\prime}{\partial H_S} & \mu\frac{\partial H_B^\prime}{\partial H_S}\frac{\partial H_S^\prime}{\partial H_S}\\
 0 & 0 & 0 & 0 & \vline & 0 & 0 & \left(\frac{\partial H_S^\prime}{\partial H_S}\right)^2
 \end{pmatrix}\,}\label{eq:16}
\end{equation}
where $\mu = \sqrt{N_\text{nn}/N_\text{n}}$.  At a fixed point,
$\mathbf{V} = \mathbf{V}^\prime \equiv \mathbf{V}^\ast$, so that
$\mathbf{V}^\ast$ is the left eigenvector with eigenvalue $b^d$ of
the extended recursion matrix evaluated at the fixed point,
$\overleftrightarrow{\mathbf{W}}^\ast$.  To evaluate $\mathbf{V}$
for an initial system away from the fixed poþnt, Eq.~\eqref{eq:15}
is iterated along the renormalization-group trajectory,
\begin{equation}\label{eq:17}
\mathbf{V} = b^{-nd} \mathbf{V}^{(n)} \cdot \overleftrightarrow{\mathbf{W}}^{(n)} \cdot \overleftrightarrow{\mathbf{W}}^{(n-1)} \cdots \overleftrightarrow{\mathbf{W}}^{(1)}\,,
\end{equation}
where $\mathbf{V}^{(n)}$ is evaluated in the system reached after
the $n$th renormalization-group step, at which
$\overleftrightarrow{\mathbf{W}}^{(n)}$ is evaluated. When the total
number of renormalization-group steps $n$ is large enough so that
the neighborhood of a fixed point is reached, $\mathbf{V}^{(n)}
\simeq \mathbf{V}^\ast$, so that $\mathbf{V}$ is evaluated to a
desired accuracy, by adjusting $n$.

From the recursion relations in Eqs.~\eqref{eq:10},\eqref{eq:10b},
the extended recursion matrix $\overleftrightarrow{\mathbf{W}}$ is
\begin{equation}
{\footnotesize
\overleftrightarrow{\mathbf{W}}=\begin{pmatrix} 4 & 2u & 0 & 0 &\vline & 4v   & {\sqrt{6}}v & \frac{3v}{2} \\
        0 & 2u & 0 & 0 & \vline &-4u^2 & \sqrt{6}u^2  & -\frac{3u^2}{2} \\
        0 & 0 & 2 + 2u & \frac{3u}{2} &\vline & 0 & 0 & 0\\
        0 & 0 & 0 & 1 & \vline &0 & 0 & 0\\ \hline
        0 & 0 & 0 & 0 &\vline &(2+2u)^2 & \sqrt{6}u\left( 1 + u \right) & \frac{3u^2}{2} \\
        0 & 0 & 0 & 0 &\vline &0 &2 + 2u &\sqrt{\frac{3}{2}}u\\
        0 & 0 & 0 & 0 &\vline &0 &0 & 1 \end{pmatrix},}\label{eq:18}
\end{equation}
where $u = \tanh 2J$, $v = 1+\text{sech}^2\, 2J$.  At the sink of
the disordered phase, $u = 0$, $v = 2$, and the left eigenvector of
$\overleftrightarrow{\mathbf{W}}^\ast$ with eigenvalue $b^d$ is
\begin{align}
\mathbf{V}^\ast =& (1,\,U=0,\,M_B=0,\,M_S=0,\nonumber\\
&\qquad
\chi_{BB}=\infty,\,\chi_{BS}=\sqrt{6},\,\chi_{SS}=1)\,.\label{eq:19}
\end{align}
The matrix multiplication of Eq.~\eqref{eq:17} mixes $\chi_{BB}$,
$\chi_{BS}$, and $\chi_{SS}$.  Since $\chi_{BB} = \infty$ at the
sink, all three susceptibilities are infinite within the disordered
phase.  In contrast, at the sink of the ordered phase, $u = 2$,
$v=1$, and the two left eigenvectors of
$\overleftrightarrow{\mathbf{W}}^\ast$ with eigenvalue $b^d$ are
\begin{align}
\mathbf{V}^\ast_\pm =& (1,\,U=1,\,M_B=\pm 2,\,M_S=\pm 1,\nonumber\\
&\qquad \chi_{BB}=0,\,\chi_{BS}=0,\,\chi_{SS}=0)\,.\label{eq:20}
\end{align}
Consequently, the susceptibilities from Eq.~\eqref{eq:17} are finite
within the ordered phase, decreasing to zero as zero temperature is
approached and increasing as $|t|^{-\gamma}$ as $T_c$ is approached
from below.  The double value in Eq.~\eqref{eq:20} reflects the
first-order phase transition along the magnetic field direction.

The infinite susceptibility in the disordered phase is directly
related to the presence of sites with arbitrarily high degree
numbers in the scale-free network, because these sites feel a very
large applied field, channeled through their many neighbors.  Except
for this feature, the critical behavior for the $p=0$ case is
similar to that of a regular lattice, which is unsurprising since
the Migdal-Kadanoff recursion relations that are exact on the
hierarchical lattice can be derived from a bond-moving approximation
applied to the square lattice.

\begin{figure}
\centering\includegraphics*[scale=1]{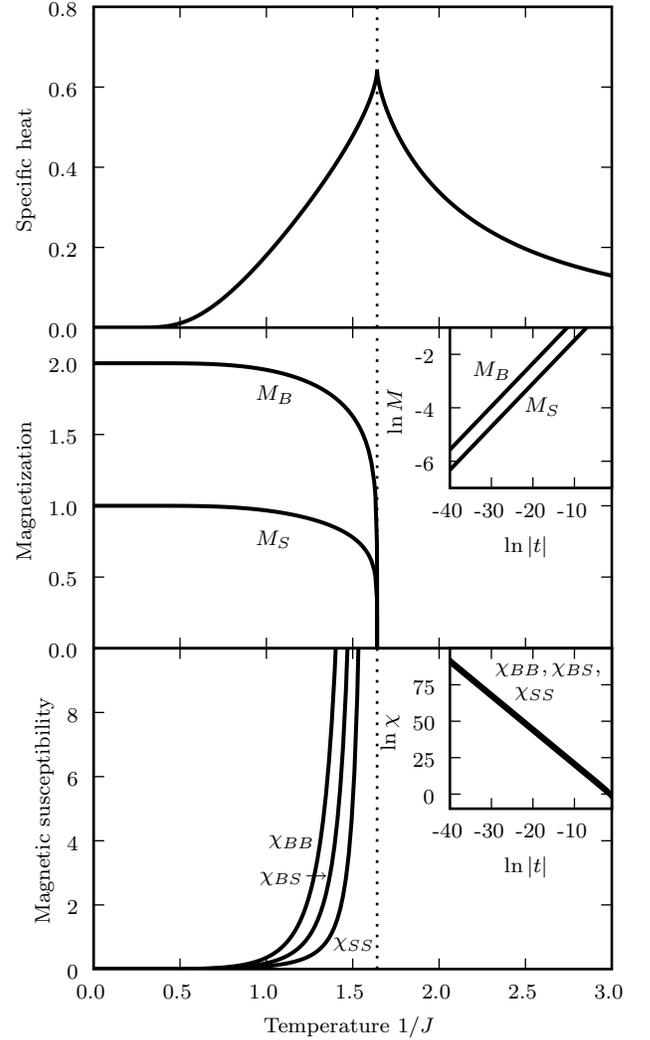}
\caption{Specific heat, magnetizations, and zero-field magnetic
susceptibilities for $p=0$, as functions of temperature $1/J$. The
dotted vertical line marks the critical temperature $T_c = 1.641$.
As insets to the magnetizations and susceptibilities, we show $\ln
M$ and $\ln \chi$ with respect to $\ln |t|$, where $t=
\frac{T-T_c}{T_c} <0$.  The linear behavior in the insets agrees
with the power-law predictions of $M_B$, $M_S \sim |t|^{0.162}$ and
$\chi_{BB}$, $\chi_{BS}$, $\chi_{SS} \sim
|t|^{-2.353}$.}\label{fig:p0dens}
\end{figure}

The $p=0$ results are in Fig.~\ref{fig:p0dens}, where the specific
heat, magnetizations, and zero-field susceptibilities are plotted as
a function of temperature.  Since the specific heat exponent is
$\alpha = -0.677$, the specific heat has a finite cusp singularity
at $T_c$.

\subsection{Critical Properties at $p=1$}

\begin{figure*}
  \centering\includegraphics*[scale=1]{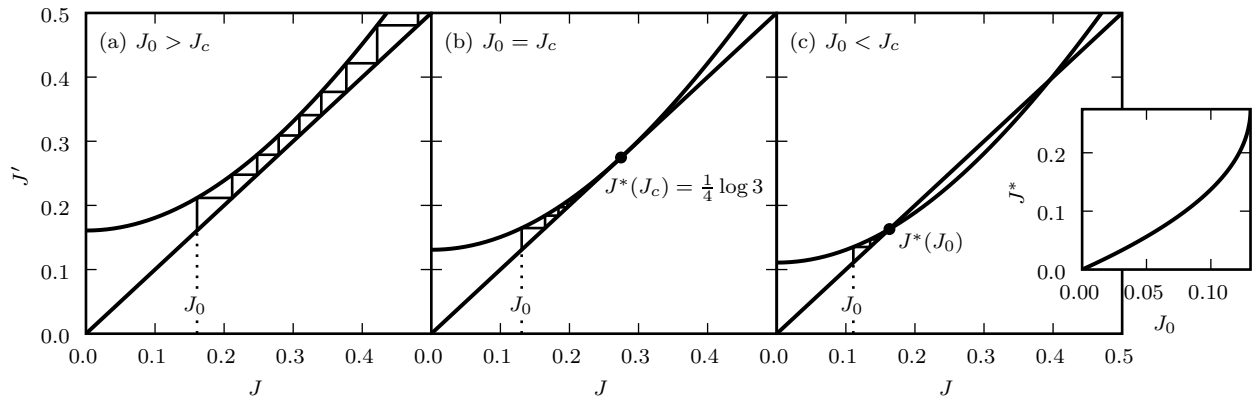} \caption{Three
    possible behaviors of the renormalization-group flows of the $p=1$
    network with uniform long-range bonds.  The curve in each diagram
    is the recursion $J^\prime(J)$ from Eq.~\eqref{eq:24}, with the
    straight line $J^\prime = J$ also drawn for reference.
    Intersections of the curve with the straight line are fixed
    points. The flows are given by the staggered line, with successive
    values of $J^\prime$ corresponding to where the staggered line
    touches the curve.  Only the dotted fixed points are physically
    accessible.  The inset on the right shows the continuous line of
    fixed points $J^\ast(J_0)$ as a function of $J_0$.}
  \label{flowfig}
\end{figure*}

For the $p=1$ lattice, the renormalization-group transformation
consists of decimating the two center sites in each connected
cluster of the type shown on the top right of Fig.~\ref{fig1}.  The
Hamiltonian now includes long-range bonds, Eq.~\eqref{eq:9aa}, and
the transformation is a mapping of the Hamiltonian $-\beta {\cal
H}(J,H_B,H_S,\{K_q\},G)$ onto a renormalized Hamiltonian
$-\beta^\prime {\cal H}^\prime
(J^\prime,H_B^\prime,H_S^\prime,\{K_q^\prime\}, G^\prime))$.  The
recursion relations are
\begin{align}
J^\prime &= \frac{1}{2}\ln \left(R_{++}R_{--}/R_{+-}^2\right) + K_1\,,\nonumber\\
H_B^\prime &= \frac{1}{2}\ln \left(R_{++}/R_{--}\right)\,,\qquad H_S^\prime = H_S\,,\nonumber\\
G^\prime &= 4 G +
\frac{1}{2}\ln\left(R_{++}R_{--}R_{+-}^2\right)\,,\nonumber\\
K_q^\prime &= K_{q+1}\,, \qquad q = 1,2,\ldots\,,\label{eq:10c}
\end{align}
where $R_{++}$, $R_{--}$, and $R_{+-}$ are as given in
Eq.~\eqref{eq:10b}.

Long-range bonds as well as nearest-neighbor bonds now contribute to
the internal energy $U$,
\begin{equation}\label{eq:23c}
U = \frac{N_\text{nn} U_\text{nn} + \sum_{q=1}^{\infty} q^{-\sigma} N_{\text{ld},q} U_{\text{ld},q}}
{N_\text{nn} + \sum_{q=1}^\infty N_{\text{ld},q}}\,,
\end{equation}
where
\begin{align}
U_\text{nn} &= \frac{1}{N_\text{nn}} \sum_{\langle i j \rangle_\text{nn}} \langle s_i s_j \rangle = \frac{1}{N_\text{nn}}\frac{\partial}{\partial J} \ln Z\,,\nonumber\\
U_{\text{ld},q} &= \frac{1}{N_{\text{ld},q}} \sum_{\langle i j \rangle_{\text{ld},q}} \langle s_i s_j \rangle = \frac{1}{N_{\text{ld},q}}\frac{\partial}{\partial K_q} \ln Z\,.\label{eq:23d}
\end{align}
Here $N_{\text{ld},q} = 4^{-q} N_\text{nn}$ is the number of
long-range bonds with $m_{ij} = q$.  Since $K_q^\prime$ does not
depend on $J$, $H_B$, or $H_S$, the thermodynamic densities and
response functions in $\mathbf{V} =
(1,U_\text{nn},M_B,M_S,\chi_{BB},\chi_{BS},\chi_{SS})$ still obey
the recursion relation in Eq.~\eqref{eq:15} with a matrix
$\overleftrightarrow{\mathbf{W}}$ of the same form as in
Eq.~\eqref{eq:16}.  The densities $U_{\text{ld},q}$, on the other
hand, have the recursion relation
\begin{align}
U_{\text{ld},1} &= b^{-d} U_\text{nn}^\prime \frac{N_\text{nn}}{N_{\text{ld},1}}\frac{\partial J^\prime}{\partial K_1} = U_\text{nn}^\prime\,,\nonumber\\
U_{\text{ld},q} &= b^{-d} U_{\text{ld},q-1}^\prime \frac{N_{\text{ld},q-1}}{N_{\text{ld},q}}\frac{\partial K_{q-1}^\prime}{\partial K_q}=U_{\text{ld},q-1}^\prime \quad (q>1)\,.
\label{eq:23f}
\end{align}
Thus $U_{\text{ld},q} = U_\text{nn}^{(q)}$, where
$U_\text{nn}^{(q)}$ is the nearest-neighor density $U_\text{nn}$ in
the system reached after $q$ renormalization-group transformations.
Thus all the long-range bond densities $U_{\text{ld},q}$ are found
by evaluating $U_\text{nn}$ along the renormalization-group
trajectory. Eq.~\eqref{eq:23c} can be rewritten as
\begin{equation}\label{eq:23g}
U = \frac{3}{4}\left(U_\text{nn} + \sum_{q=1}^\infty q^{-\sigma} 4^{-q} U_\text{nn}^{(q)}\right)\,,
\end{equation}
where we have also used $N_\text{nn} + \sum_{q=1}^\infty
N_{\text{ld},q} = \frac{4}{3} N_\text{nn}$. From Eq.~\eqref{eq:23g}
and the recursion relation for $U_\text{nn}$, the leading
singularity in $U_\text{nn}$ is also the leading singularity in $U$.
It is sufficient to calculate the singular behavior of $U_\text{nn}$
to obtain the critical properties of $U$ and of the specific heat
$C$.

\subsubsection{Long-distance bonds with uniform interaction strengths}

We first consider the case with no distance dependence in the
strengths of the long-range bonds, $\sigma = 0$.  Here $K_q = J_0$
for all $q$ and after any number of renormalization-group
transformations, where $J_0$ is the value of $J$ in the original
system. The recursion relation for $J$ in the closed subspace $H_B =
H_S = 0$ is
\begin{equation}\label{eq:24}
J^\prime = J_0 + \ln(\cosh 2J)\,.
\end{equation}
There are three types of behavior possible for the
renormalization-group flows, as illustrated in Fig.~\ref{flowfig}.
For $J_0$ greater than a critical value $J_c$
(Fig.~\ref{flowfig}(a)), the flows go to the ordered phase sink
$J^\ast = \infty$.  For $J_0 \le J_c$ (Fig.~\ref{flowfig}(b,c)) the
flows go to a continuous line of fixed points $J^\ast(J_0)$, with a
distinct fixed point for each starting interaction $J_0$. When $J_0
= J_c$ exactly, the $J^\prime(J)$ curve touches tangentially the
straight line $J^\prime = J$ at $J^\ast(J_c)$, as shown in
Fig.~\ref{flowfig}(b). This fact allows us to solve for
$J^\ast(J_c)$ and $J_c$ exactly:
\begin{equation}\label{eq:25}
J^\ast(J_c) = \frac{1}{4}\ln 3\,, \qquad J_c = \ln
\frac{3^{3/4}}{2}\,.
\end{equation}

Thus the system is conventionally ordered below the critical
temperature $T_c = 1/J_c = 7.645$.  To understand the novel
high-temperature phase above $T_c$, we look at the recursion matrix
$\overleftrightarrow{\mathbf{W}}^\ast$ evaluated along the line of
fixed points, $J^\ast(J_0)$ for $J_0 \le J_c$.  The form of the
matrix is as in Eq.~\eqref{eq:18}, with $u=\tanh 2J^\ast(J_0)$ and
$v = 1+ \text{sech}^2\, 2J^\ast(J_0)$.  Since $J^\ast(J_0)$ has the
maximum value of $(\ln 3)/4 = 0.275$ for $J_0 = J_c$ and tends to
zero as $J_0$ increases, $0 \le u \le 1/2$, $7/4 \le v \le 2$. The
left eigenvector of $\overleftrightarrow{\mathbf{W}}^\ast$ with
eigenvalue $b^d$ is
\begin{align}
\mathbf{V}^\ast =& (1,\,U_\text{nn}=\frac{u}{2-u},\,M_B=0,\,M_S=0,\nonumber\\
&\qquad
\chi_{BB}=\infty,\,\chi_{BS}=\infty,\,\chi_{SS}=\infty)\,.\label{eq:26}
\end{align}
It follows that, in the high-temperature phase, $M_B = M_S = 0$ and
that the susceptibilities $\chi_{BB}$, $\chi_{BS}$, $\chi_{SS}$ are
infinite.  Because the renormalization-group flows go to a line of
fixed points ending at the critical point $J^\ast(J_c)$, the
correlation length is infinite throughout the phase and the
correlations have power-law decay, characteristics which are
typically seen just at $T=T_c$.  (In contrast, the low-temperature
ordered phase has the usual exponential decay of correlations.) This
type of behavior, with a transition between phases with finite and
infinite correlation lengths, was first seen in the
Berezinskii-Kosterlitz-Thouless phase
transition~\cite{Berezinskii,KosterlitzThouless}, though with an
important difference:  There the algebraic order was in the
low-temperature phase, while here it is the high-temperature phase
that has this feature.

We now turn to the critical behavior of the system in the ordered
phase, as $T \to T_c$ from below. For small negative $t =
(T-T_c)/T_c = (J_c-J_0)/J_0$, we have $J_0 = J_c + \delta$, where
$\delta = J_c |t|$.  As can be seen from Fig.(6a), a
renormalization-group flow starting at $J_0$ spends a large number
of iterations in the vicinity of $J^\ast(J_c) = (\ln 3)/4$, before
escaping to the ordered phase sink at $J^\ast = \infty$.  If $n_0$
is the number of iterations initially required to get $J$ close to
$J^\ast(J_c)$ and $n^\ast$ is the number of iterations where $J
\approx J^\ast(J_c)$, then as $\delta \to 0$, $n_0$ remains
constant, while $n^\ast \to \infty$.  The dependence of $n^\ast$ on
$\delta$ (and hence on $|t|$) determines the critical singularities.
For a typical critical point, $n^\ast \sim (\ln\delta)/(y_T \ln b)$.
However, in our case, at $J^\ast(J_c)$ the eigenvalue exponent $y_T
= 0$, and it turns out that $n^\ast \sim \delta^{-1/2}$.  We show
this as follows: After $n_0$ iterations, the flow is at $J$ near
$J^\ast(J_c)$, with $J < J^\ast(J_c)$.  It then takes $n^\ast/2$
iterations to get $J$ almost exactly at $J^\ast(J_c)$, and another
$n^\ast/2$ iterations to get $J$ a significant distance away from
$J^\ast(J_c)$, namely to $J - J_c \sim \text{O}(1)$.  Considering
the latter half of this flow, we expand the recursion relation for
$J$, Eq.~\eqref{eq:24}, around $J^\ast(J_c)$,
\begin{align}
J^\prime - J^\ast(J_c) =&~ \delta + (J-J^\ast(J_c)) + \frac{3}{2}(J-J^\ast(J_c))^2\nonumber\\
 &\qquad -(J-J^\ast(J_c))^3 + \cdots\,.\label{eq:27}
\end{align}
Starting with $J = J^\ast(J_c)$, from Eq.~\eqref{eq:27}, we obtain
series expressions for $J^{(i)}$, the interaction after $i$
renormalization-group steps:
\begin{align}
J^{(1)} - J^\ast(J_c) &= \delta\nonumber\\
J^{(2)} - J^\ast(J_c) &= 2\delta + \frac{3}{2}\delta^2 -\delta^3 + \cdots\nonumber\\
&.\:.\:.\nonumber\\
J^{(n)} - J^\ast(J_c) &= n\delta + \frac{1}{4}(n-1)n(2n-1)\delta^2\nonumber\\
+ \frac{1}{80}&(n-3)(n-1)n(24n^2-29n+2)\delta^3 +
\cdots\label{eq:28}
\end{align}
For $n \ll \delta^{-1/2}$, the first term in the series for $J^{(n)}
- J^\ast(J_c)$ is dominant, and the distance increases very slowly
as $J^{(n)} - J^\ast(J_c) \simeq n\delta$.  For large $n$, the $k$th
term in the series $\sim n^{2k-1} \delta^k$. Thus, when $n$ is of
the order $\delta^{-1/2}$, $J^{(n)} - J^\ast(J_c)$ begins to
increase significantly.  From this we can deduce that $n^\ast$ scales
like $\delta^{-1/2}$.

We can now proceed to find the critical behaviors for the
correlation length, thermodynamic densities, and response functions.
By iterating the recursion relation for the correlation length, $\xi
= b \xi^\prime$, $\xi = b^{n^\ast + n_0} \xi^{(n+n_0)}$, where
$\xi^{(n)}$ is the correlation length after $n$
renormalization-group steps.  The singularity in $\xi$ as $\delta
\to 0$ comes from the $b^{n^\ast}$ factor,
\begin{equation}\label{eq:29}
\xi \sim b^{n^\ast} \sim e^{\frac{C\ln 2}{\sqrt{\delta}}} =
e^{\frac{A}{\sqrt{|t|}}}\,,
\end{equation}
where $n^\ast \approx C \delta^{-1/2}$ for some constant $C$, and $A =
C/\sqrt{J_c}$.

From Eqs.~\eqref{eq:16},\eqref{eq:17}, we extract the critical
behaviors of the internal energy, magnetizations, and
susceptibilities:  The nearest-neighbor contribution to the internal
energy $U_\text{nn}$ transforms as
\begin{equation}\label{eq:30}
U_\text{nn} = b^{-d} \frac{\partial G^\prime}{\partial J} + b^{-d} U_\text{nn}^\prime \frac{\partial J^\prime}{\partial J}\,.
\end{equation}
Since $\partial G^\prime / \partial J$ is analytic, the singularity
of $U_\text{nn}$ must reside in $U_{\text{nn}_{\text{\scriptsize
sing}}} = b^{-d} U_\text{nn}^\prime \partial J^\prime / \partial J$.
Iterating over $n_0+n^\ast$ renormalization-group steps,
\begin{align}
U_{\text{nn}_{\text{\scriptsize sing}}} &= b^{-(n_0+n^\ast)d} U_\text{nn}^{(n_0+n^\ast)} \prod_{i=1}^{n_0+n^\ast} \left.\frac{\partial J^\prime}{\partial J}\right|_{J=J^{(i)}}\nonumber\\
&\simeq b^{-n^\ast d}\left[ b^{-n_0 d} U_\text{nn}^{(n_0+n^\ast)}
\prod_{i=1}^{n_0} \left.\frac{\partial J^\prime}{\partial
J}\right|_{J=J^{(i)}}\right] \label{eq:31}\,,
\end{align}
where we have used the fact that $\partial J^\prime / \partial J
\simeq 1$ for the $n^\ast$ iterations during which $J^{(i)} \simeq
J^\ast(J_c)$.  After $n+n^\ast$ iterations the system has flowed
away from criticality.  The singular dependence comes from the
$b^{-n^\ast d}$ factor,
\begin{equation}\label{eq:32}
U_{\text{nn}_{\text{\scriptsize sing}}} \sim b^{-n^\ast d} \sim
e^{-\frac{dA}{\sqrt{|t|}}}\,.
\end{equation}
Thus the singular part of the specific heat is
\begin{equation}\label{eq:32b}
C_{\text{\scriptsize sing}} \sim |t|^{-3/2}
e^{-\frac{dA}{\sqrt{|t|}}}.
\end{equation}

The magnetizations $M_B$ and $M_S$ recur as
\begin{align}
(M_B,M_S) &= b^{-d} (M_B^\prime,M_S^\prime)\begin{pmatrix} 2+2u &
\frac{3}{2}u \\ 0 & 1 \end{pmatrix}\,,\label{eq:33}
\end{align}
where $u = \tanh 2J$.  Iterating over $n_0 + n^\ast$
renormalization-group steps,
\begin{align}
&(M_B, M_S) \nonumber\\
&\:\simeq b^{-(n_0+n^\ast)d}
\left((M_B^{(n_0+n^\ast)},M_S^{(n_0+n^\ast)})\cdot\mathbf{v}\right)
b^{n^\ast y_H} \mathbf{v} \cdot \mathbf{R}\,.\label{eq:34}
\end{align}
Here $b^{y_H} = 2+ 2 \tanh 2J^\ast(J_c) = 3$ is the largest
eigenvalue of the $2\times 2$ derivative matrix in Eq.~\eqref{eq:33}
evaluated at $J^\ast(J_c)$, $\mathbf{v}$ is the corresponding
normalized (to unity) eigenvector, and $\mathbf{R}$ is the product
of the derivative matrices of the first $n_0$ iterations.  The
singular behavior comes from the factor $b^{-n^\ast (d-y_H)}$,
\begin{equation}\label{eq:35}
M_B,\,M_S \sim b^{-n^\ast(d-y_H)} \sim
e^{-\frac{(d-y_H)A}{\sqrt{|t|}}}\,.
\end{equation}
Since $y_H = 1.585$, the magnetizations decrease exponentially to
zero as $|t| \to 0$.

The susceptibilities $\chi_{BB}$, $\chi_{BS}$, and $\chi_{SS}$ recur
as
\begin{align}
&{\small
(\chi_{BB},\chi_{BS},\chi_{SS})}\nonumber\\
&{\small
= b^{-d} (\chi_{BB}^\prime,\chi_{BS}^\prime,\chi_{SS}^\prime)\begin{pmatrix} (2+2u)^2 & \sqrt{6}u(1+u) & \frac{3u^2}{2}\\
0 & 2+2u & \sqrt{\frac{3}{2}}u\\ 0 & 0 & 1\end{pmatrix}}\nonumber\\
&{\small
\qquad + b^{-d} (G^\prime, U_\text{nn}^\prime) \begin{pmatrix} 4v   & {\sqrt{6}}v & \frac{3v}{2} \\
-4u^2 & \sqrt{6}u^2  & -\frac{3u^2}{2}\end{pmatrix}\,,}\label{eq:36}
\end{align}
where $v=1+\text{sech}^2\,2J$.  Since there is no singular behavior
in $G^\prime$ and $U^\prime_{\text{nn}_{\text{\scriptsize sing}}}
\to 0$ as $|t| \to 0$, only the first term in Eq.~\eqref{eq:36}
contributes to the divergent singularity of the susceptibilities.
Iterating over $n_0 + n^\ast$ steps,
\begin{align}
&(\chi_{BB},\chi_{BS},\chi_{SS})_\text{sing} \sim b^{-(n_0+n^\ast)d}\nonumber\\
&\:\left((\chi_{BB}^{(n_0+n^\ast)},\chi_{BS}^{(n_0+n^\ast)},\chi_{SS}^{(n_0+n^\ast)})\cdot
\mathbf{v}\right) b^{2n^\ast y_H}
\mathbf{v}\cdot\mathbf{R}\,,\label{eq:37}
\end{align}
where $b^{2y_H} = (2+2\tanh2J^\ast(J_c))^2$ is the largest
eigenvalue of the $3\times 3$ derivative matrix in Eq.~\eqref{eq:36}
evaluated at $J^\ast(J_c)$, $\mathbf{v}$ the corresponding
normalized eigenvector, and $\mathbf{R}$ the product of the
derivative matrices for the first $n_0$ steps.  The singularity in
the susceptibilities is given by
\begin{equation}\label{eq:38}
{\chi_{BB}}_\text{sing},\,{\chi_{BS}}_\text{sing},\,{\chi_{SS}}_\text{sing}
\sim b^{n^\ast(2y_H -d)} \sim e^{\frac{(2y_H-d)A}{\sqrt{|t|}}}\,.
\end{equation}

\begin{figure}
\centering\includegraphics*[scale=1]{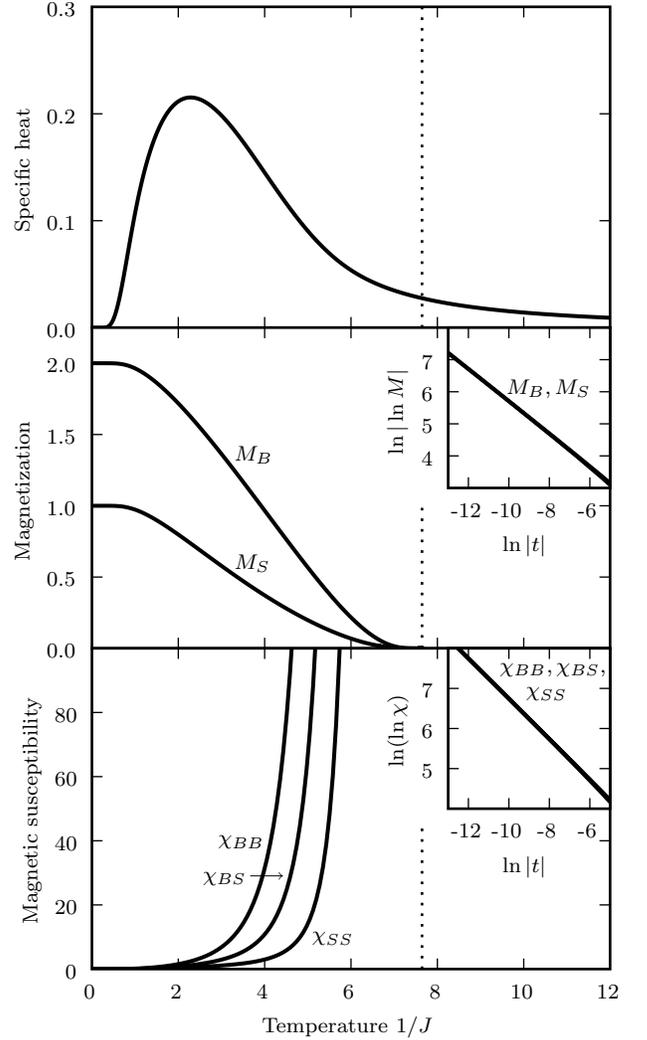}
\caption{Specific heat, magnetizations, and magnetic
susceptibilities for $p=1$ long-range bonds with uniform interaction
strengths ($\sigma = 0$), as a function of temperature $1/J$. The
dotted vertical line marks the critical temperature $T_c = 7.645$.
The insets to the magnetization and susceptibility graphs show
$\ln|\ln M|$ and $\ln(\ln \chi)$ versus $\ln |t|$, where $t=
\frac{T-T_c}{T_c} <0$. The linear behavior in the insets agrees with
the exponential scaling predictions of $M_B$, $M_S \sim
e^{-C/\sqrt{|t|}}$ and $\chi_{BB}$, $\chi_{BS}$, $\chi_{SS} \sim
e^{D/\sqrt{|t|}}$ with positive constants $C$ and $D$
.}\label{fig:p1dens}
\end{figure}

We illustrate these results in Fig.~\ref{fig:p1dens}, plotting the
specific heat, magnetizations, and zero-field susceptibilities as a
function of temperature.  The essential singularity in the specific
heat (Eq.~\eqref{eq:32b}) is invisible in the plot, the function and
all its derivatives being continuous at $T_c$, with the rounded
analytic peak occurring in the phase opposite to the algebraic
phase, namely in the ordered phase at lower temperature. This
behavior of the specific heat also occurs in the XY model undergoing
a Berezinskii-Kosterlitz-Thouless phase transition, as seen in
Fig.~5 of Ref.\cite{BerkerNelson}.  In the latter case, opposite to
the algebraic phase, the phase in which the rounded analytic peak
occurs is the disordered phase at higher temperature.  In the XY
model, the physical meaning of the high-temperature rounded peak is
the onset of short-range order within the disordered phase. In our
current system, the physical meaning of the low-temperature rounded
peak is the saturation of long-range order that occurs unusually
away from criticality, due to the essential critical singularity of
the magnetization, which corresponds to a critical exponent $\beta =
\infty$ and the unusual flat onset of the magnetization, as seen in
Fig.~\ref{fig:p1dens}.

\subsubsection{Long-distance bonds with decaying interaction strengths}

\begin{figure}
\includegraphics*[scale=1]{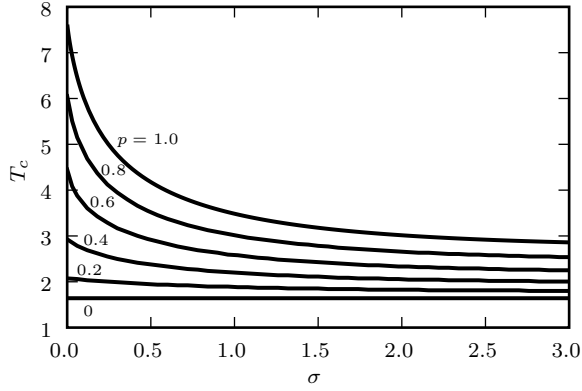}
\caption{Critical temperature $T_c = 1/J_c$ for various $p$ as a
  function of the exponent $\sigma$ describing the decay of the
  long-range bond interaction strengths.  The curves for $0 < p < 1$
  were calculated using the techniques in Section III.C. }\label{fig:p1tc}
\end{figure}

For $\sigma > 0$, the long-range bond strengths $K_q = J_0
q^{-\sigma}$ and thus at the $n$th renormalization-group step
$K_1^{(n)} = J_0 (n+1)^{-\sigma}$.  The interaction strength
$J^{(n)}$ in the closed subspace $H_B = H_S = 0$ is given by the
recursion relation
\begin{equation}
J^{(n)} = J_0n^{-\sigma} + \ln(\cosh 2J^{(n-1)}) \label{eq:39}\,,
\end{equation}
where $J^{(0)} = J_0$.  The critical temperature $T_c$ now depends on
the exponent $\sigma$, as shown in top curve of Fig.~\ref{fig:p1tc},
having the maximum value of $T_c = 7.645$ at $\sigma = 0$ and
decreasing with increasing $\sigma$ (to $T_c = 2.744$ at $\sigma =
\infty$, where the system reduces to a nearest-neighbor,
next-nearest-neighbor model).

When the number of renormalization-group steps $n \to \infty$, the
$J_0 n^{-\sigma}$ term in Eq.~\eqref{eq:39} goes to zero, so that
the fixed points of the renormalization-group transformation are
those of the $p=0$ case analyzed in Sec. III.A.  Thus for
temperatures close enough to $T_c$, satisfying $|t| \ll \tau$ for
some crossover value $\tau$, we expect to observe the $p=0$ critical
behavior. However, the width $\tau$ of the critical region varies
with $\sigma$, becoming extremely narrow as $\sigma \to 0$.  For a
thermodynamic quantity scaling as $|t|^x$ inside the critical region
($x$ being one of the $p=0$ exponents), the general scaling behavior
for small $|t|$ not necessarily in the critical region is
$|t|^{x+f_x(t)}$, where $|f_x(t)| \ll |x|$ when $|t| \ll \tau$, and
the form of $f_x(t)$ may depend on $\sigma$.   In the following, we
derive the leading order contribution to $f_x(t)$ for the various
physical properties of the system, also determining the size of the
critical region $\tau$.

If the system is at its critical temperature, $J_0 = J_c$, the
interaction strength under repeated renormalization-group
iterations, $J^{(n)}$ for $n \to \infty$, goes to the $p=0$ critical
fixed point, which we will label $J_{c0}$ and whose value is given
by Eq.~\eqref{eq:11}.  Let us denote this renormalization-group flow
as $J_c^{(n)}$, so that $J_c^{(0)} = J_c$ and $\lim_{n\to\infty}
J_c^{(n)} = J_{c0}$.  Now if we start instead at a temperature very
close to critical, $J_0 = J_c - J_c t$ for small $|t|$, $J^{(n)}$
stays near $J^{(n)}_c$ for a large number of iterations $n^\ast$,
before veering off to either the ordered or disordered sink.  The
dependence of $n^\ast$ on $|t|$ is the key to the crossover behavior
of the system.  The difference $J^{(n)} - J_c^{(n)}$ satisfies the
recursion relation
\begin{equation}\label{eq:40}
J^{(n+1)} - J_c^{(n+1)} = b^{y_T(n)} (J^{(n)} - J_c^{(n)})\,,
\end{equation}
where
\begin{equation}\label{eq:41}
b^{y_T(n)} = \left.\frac{\partial J^{(n+1)}}{\partial
J^{(n)}}\right|_{J^{(n)}=J_c^{(n)}} = 2 \tanh 2J_c^{(n)}\,.
\end{equation}
Iterating Eq.~\eqref{eq:40},
\begin{align}
J^{(n+1)} - J_c^{(n+1)} &= b^{\sum_{k=0}^{n} y_T(k)}(J_0 - J_c)\nonumber\\
&= b^{\sum_{k=0}^{n} y_T(k)} (-J_c t)\label{eq:42}\,.
\end{align}
Since $J^{(n^\ast)} - J_c^{(n^\ast)} \sim \text{O}(1)$,
\begin{equation}
b^{\sum_{k=0}^{n^\ast} y_T(k)} \sim |t|^{-1}\ \quad \text{and} \quad
\sum_{k=0}^{n^\ast} y_T(k) \sim -\frac{\ln |t|}{\ln
b}\,.\label{eq:43}
\end{equation}
In order to find $n^\ast$, we need to determine $y_T(n)$.  From the
fact that $\lim_{n\to\infty} J_c^{(n)} = J_{c0}$ and the recursion
relation in Eq.~\eqref{eq:39}, we consider for $J_c^{(n)}$ the large
$n$ form of
\begin{equation}\label{eq:44}
J_c^{(n)} = J_{c0} - B n^{-\sigma} + \cdots\,.
\end{equation}
Substitution into Eq.~\eqref{eq:39} yields
\begin{equation}\label{eq:46}
B = \frac{J_0}{2\tanh 2J_{c0}-1}\,.
\end{equation}
Eqs.~\eqref{eq:44},\eqref{eq:46} can also be obtained by expanding
the recursion relation around $J_{c0}$,
\begin{equation}
J_c^{(n)}-J_{c0} = J_0n^{-\sigma} + 2\tanh
2J_{c0}(J_c^{(n-1)}-J_{c0})\label{eq:46b}\,,
\end{equation}
and summing the series derived from iterating Eq.~\eqref{eq:46b}.
Substituting into Eq.~\eqref{eq:41},
\begin{equation}
y_T(n) = y_{T0} - C n^{-\sigma}+\cdots\,,\label{eq:47}
\end{equation}
where $y_{T0} = \ln(2\tanh 2J_{c0})/\ln b = 0.747$ is the $p=0$
thermal eigenvalue exponent and $C = 1.498 J_0$.  For use below, we
also deduce the magnetic exponents $y_H(n)$,
\begin{align}
b^{y_H(n)} &= \left.\frac{\partial H_B^{(n+1)}}{H_B^{(n)}}\right|_{J^{(n)}=J_c^{(n)},H_B^{(n)}=H_S^{(n)}=0}\nonumber\\
&= 2+2\tanh 2J^{(n)}_c,\nonumber\\
y_H(n) &= y_{H0} - D n^{-\sigma}+\cdots\,.\label{eq:49}
\end{align}
where $y_{H0} = \ln(2+2\tanh 2J_{c0})/\ln b = 1.879$ is the $p=0$
magnetic eigenvalue exponent and  $D = 0.683 J_0$.

From Eq.~\eqref{eq:47}, we evaluate $\sum_{k=0}^{n^\ast} y_T(k)$ for
large $n^\ast$,
\begin{equation}\label{eq:52}
\sum_{k=0}^{n^\ast} y_T(k) \simeq \begin{cases} n^\ast y_{T0} - \frac{C {n^\ast}^{1-\sigma}}{1-\sigma}\,, & 0 < \sigma < 1\,,\\
n^\ast y_{T0} - C \ln n^\ast, & \sigma =1\,,\\
n^\ast y_{T0} - C \zeta(\sigma)\,, & \sigma > 1\,.
                                   \end{cases}
\end{equation}
For $\sigma \ge 1$, the $n^\ast y_{T0}$ term is clearly dominant for
large $n^\ast$, so that, from Eq.~\eqref{eq:43},
\begin{equation}\label{eq:53}
n^\ast \simeq -\frac{1}{y_{T0}} \frac{\ln{|t|}}{\ln b} \equiv
n^\ast_0 \qquad (\sigma \ge 1)\,.
\end{equation}
This expression for $n^\ast$ leads to the same critical exponents we
found in the $p=0$ case.  On the other hand, for the slow decay of
$0 < \sigma < 1$, Eq.~\eqref{eq:43} becomes
\begin{equation}\label{eq:54}
n^\ast y_{T0} - \frac{C {n^\ast}^{1-\sigma}}{1-\sigma} \simeq
-\frac{\ln |t|}{\ln b}\,.
\end{equation}
Writing $n^\ast = n^\ast_0 + \delta n$, the leading order
contribution to $\delta n$ is found,
\begin{equation}\label{eq:55}
n^\ast = n^\ast_0 + \frac{C {n^\ast_0}^{1-\sigma}}{(1-\sigma)y_{T0}}
+ \cdots \qquad (0<\sigma <1)\,.
\end{equation}

This expression for $n^\ast$ when $0<\sigma < 1$ yields the
leading-order corrections to $p=0$ in the critical behaviors of the
correlation length, internal energy, specific heat, magnetizations,
and susceptibilities:
\begin{align}
\xi &\sim b^{n^\ast} \sim
|t|^{-\frac{1}{y_{T0}}-\frac{C}{(1-\sigma)y_{T0}^{2-\sigma}}\left(-\frac{\ln
|t|}{\ln b}\right)^{-\sigma}}\,,\nonumber\\
U_\text{sing} &\sim b^{-n^\ast d + \sum_{k=0}^{n^\ast} y_T(k)} \nonumber\\
& \sim
|t|^{\frac{d-y_{T0}}{y_{T0}}+\frac{dC}{(1-\sigma)y_{T0}^{2-\sigma}}\left(-\frac{\ln
|t|}{\ln b}\right)^{-\sigma}}\,,\nonumber\\
C_\text{sing}&\sim
|t|^{\frac{d-2y_{T0}}{y_{T0}}+\frac{dC}{(1-\sigma)y_{T0}^{2-\sigma}}\left(-\frac{\ln
|t|}{\ln b}\right)^{-\sigma}}\,,\nonumber\\
M_B,\,M_S &\sim b^{-n^\ast d + \sum_{k=0}^{n^\ast} y_H(k)}\nonumber\\
& \sim |t|^{\frac{d-y_{H0}}{y_{T0}}+\frac{(d-y_{H0})C+
y_{T0}D}{(1-\sigma)y_{T0}^{2-\sigma}}\left(-\frac{\ln |t|}{\ln
b}\right)^{-\sigma}}\,,\nonumber\\
\chi_{BB},\,\chi_{BS},\,\chi_{SS} &\sim b^{-n^\ast d + 2\sum_{k=0}^{n^\ast} y_H(k)}\nonumber\\
&\sim |t|^{\frac{d-2y_{H0}}{y_{T0}}+\frac{(d-2y_{H0})C+2
y_{T0}D}{(1-\sigma)y_{T0}^{2-\sigma}}\left(-\frac{\ln |t|}{\ln
b}\right)^{-\sigma}}\,. \label{eq:60}
\end{align}
All the critical behavior expressions in Eqs.~\eqref{eq:60} have the
form $|t|^{x + \frac{E}{(1-\sigma)y_{T0}^{2-\sigma}}\left(-\frac{\ln
|t|}{\ln b}\right)^{-\sigma}}$, where $x$ is the appropriate $p=0$
exponent and $E$ is a non-universal (i.e., $J_0$-dependent) constant
$\sim \text{O(1)}$. For temperatures $|t| < \tau$, the leading-order
correction term in the exponent should be negligible,
\begin{equation}\label{eq:61}
\frac{1}{(1-\sigma)y_{T0}^{2-\sigma}}\left(-\frac{\ln |t|}{\ln
b}\right)^{-\sigma} \lesssim \epsilon\,,
\end{equation}
for some small quantity $\epsilon$, giving an estimate for $\tau$ as $\sigma \to 0$,
\begin{equation}\label{eq:62}
\tau \approx b^{-
\left(\epsilon(1-\sigma)y_{T0}^{2-\sigma}\right)^{-1/\sigma}}\,.
\end{equation}
With decreasing $\sigma$ the critical region $\tau$ becomes rapidly
infinitesimal.  For example, with $\sigma = 1/2$ and $\epsilon = 10^{-1}$,
$\tau \approx 10^{-289}$.

\begin{figure}[t]
\centering\includegraphics*[scale=1.0]{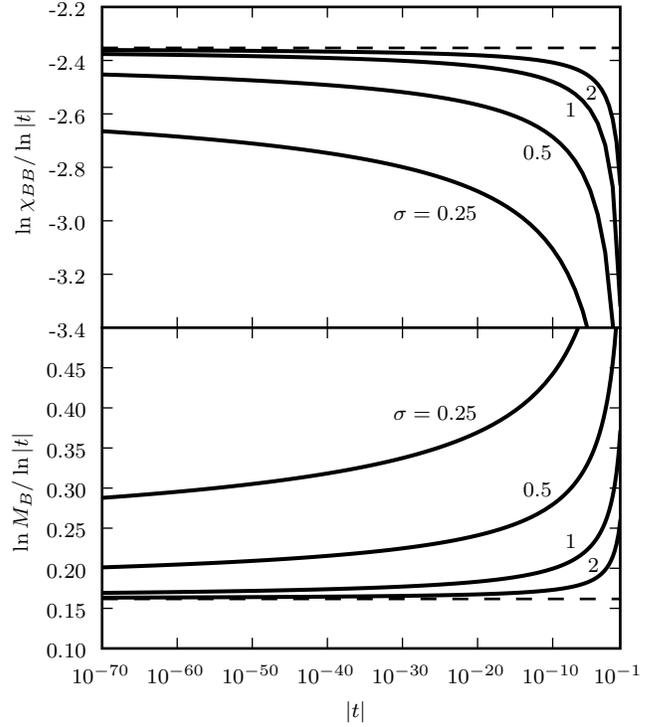} \caption{The
calculated effective exponent of the magnetization $M_B$ and
magnetic susceptibility $\chi_{BB}$, as a function of $|t|$ for $t =
\frac{T-T_c}{T_c} < 0$ and $p=1$.  Curves for several values of
$\sigma$, the exponent for the decay of the long-range bond
strengths, are shown.  The horizontal dashed line in the upper graph
corresponds to the actual critical exponent for the susceptibility,
$-\gamma = -(2y_{H0}-d)/y_{T0} = -2.353$, while the dashed line in
the lower graph corresponds to the actual magnetization exponent,
$\beta = (d-y_{T0})/y_{T0} = 0.162$.}\label{fig:p1exp}
\end{figure}

\begin{figure}[t]
  \centering\includegraphics*[scale=1.0]{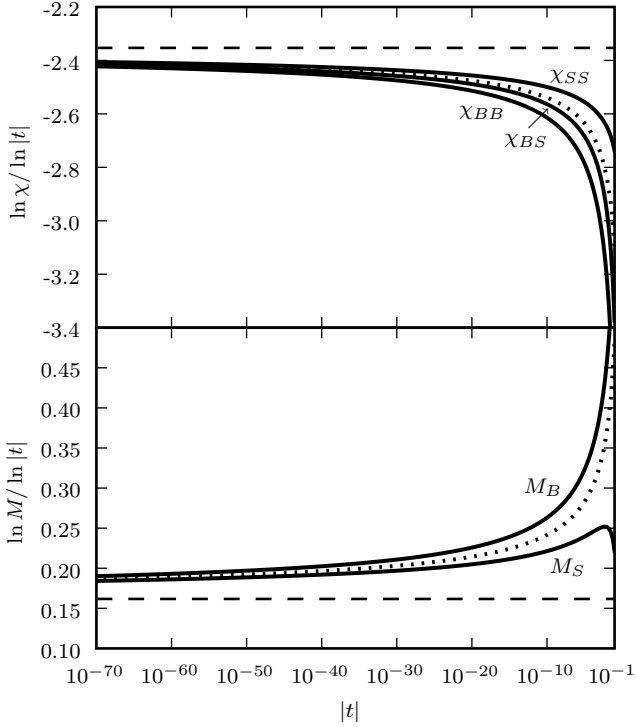}
  \caption{The effective exponents of the magnetizations $M_B$ and
    $M_S$, and susceptibilities $\chi_{BB}$, $\chi_{BS}$, and
    $\chi_{SS}$, for $p=1$, $\sigma = 0.6$, plotted as a function of
    $|t|$ for $t = \frac{T-T_c}{T_c} < 0$.  The dotted curves in the
    figures are the first-correction-to-scaling predictions for the
    magnetization and susceptibility from Eq.~\eqref{eq:60}.  The
    horizontal dashed line in the upper graph corresponds to the $p=0$
    critical exponent for the susceptibility, $-\gamma =
    -(2y_{H0}-d)/y_{T0} = -2.353$, while the dashed line in the
    lower graph corresponds to the $p=0$ magnetization exponent,
    $\beta = (d-y_{T0})/y_{T0} = 0.162$.}\label{fig:p1exp2}
\end{figure}

The above corrections to critical behavior are illustrated in
Fig.~\ref{fig:p1exp}, where we plot numerically calculated effective
exponents $\ln M_{BB} / \ln|t|$ and $\ln \chi_{BB} / \ln|t|$ as a
function of $|t|$ for several values of $\sigma$.  It is clear that
for $\sigma \ge 1$, the effective exponents quickly converge to the
horizontal lines showing the actual asymptotic exponents. The
convergence when $\sigma < 1$ is much slower, due to the $|t|^{
  \frac{E}{(1-\sigma)y_{T0}^{2-\sigma}}\left(-\frac{\ln |t|}{\ln
      b}\right)^{-\sigma}}$ correction to asymptotic universal
critical behavior.  In Fig.~\ref{fig:p1exp2} we explicitly show for
the case $\sigma = 0.6$ the magnetizations and susceptibilities
asymptotically approaching the scaling forms of Eq.~\eqref{eq:60} for
small $|t|$.

\subsection{Critical Properties of the System with Long-Range Quenched Randomness, $0<p<1$}

\subsubsection{Exact renormalization-group transformation for quenched probability distributions}

\begin{figure}[t]
\includegraphics[scale=0.5]{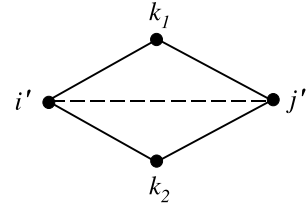}
\caption{Cluster with quenched randomness on which the decimation
  transformation of Eq.~\eqref{eq:10d} is applied.}\label{fig:cluster}
\end{figure}

When $0 < p < 1$, there is long-range quenched randomness in the
network, and the renormalized system will have an inhomogenous
distribution of all interaction constants.  The
renormalization-group transformation needs be formulated in terms of
quenched probability distributions~\cite{AndelmanBerker}.  First
consider the decimation transformation effected on the cluster of
Fig.~\ref{fig:cluster}, with nonuniform interaction constants. The
recursion relations for ${J^\prime}_{i^\prime j^\prime}$,
${H_B^\prime}_{i^\prime j^\prime}$, $H_S^\prime$, and
$G^\prime_{i^\prime j^\prime}$ are the locally differentiated
versions of Eq.~\eqref{eq:10c},
\begin{align}
J^\prime(i'j') &= J^\prime(i'k_1j') + J^\prime(i'k_2j') + K_1(i'j')\,,\nonumber\\
&J^\prime(i'k_1j') = \frac{1}{4}\ln \left(R_{++}R_{--}/R_{+-}^2\right)_{(i'k_1j')}\,,\nonumber\\
H_B^\prime(i'j') &= H_B^\prime(i'k_1j') + H_B^\prime(i'k_2j'),\quad H_S^\prime(i') = H_S(i')\,,\nonumber\\
&H_B^\prime(i'k_1j') = \frac{1}{4}\ln \left(R_{++}/R_{--}\right)_{(i'k_1j')}\,,\nonumber\\
G^\prime(i'j') &= G^\prime(i'k_1j') + G^\prime(i'k_2j')\,,\nonumber\\
&G^\prime(i'k_1j') =\frac{1}{4}\ln\left(R_{++}R_{--}R_{+-}^2\right)_{(i'k_1j')}\,,\nonumber\\
K_q^\prime &= K_{q+1}\,, \qquad q =
1,2,\ldots\,,\label{eq:10d}
\end{align}
where $R_{++}$, $R_{--}$, and $R_{+-}$ along path $(i'kj')$ are
given by the locally differentiated versions of Eq.~\eqref{eq:10b},
\begin{align}
R_{++} &= x_1 x_2 y_1 y_2 z + x_1^{-1} x_2^{-1} z^{-1}\,,\nonumber\\
R_{--} &= x_1^{-1} x_2^{-1} z + x_1 x_2 y_1^{-1} y_2^{-1} z^{-1}\,,\nonumber\\
R_{+-} &= x_1 x_2^{-1} y_1 z + x_1^{-1} x_2 y_2^{-1} z^{-1}\,,\nonumber\\
&x_1 = e^{J(i^\prime k)},\, x_2 = e^{J(k j^\prime)}\,,\nonumber\\
&y_1 = e^{2H_B(i^\prime k)},\, y_2 = e^{2H_B(k j^\prime)},\,
z=e^{H_S(k)}. \label{eq:10e}
\end{align}
If there is no long-range bond connecting $i^\prime$ and $j^\prime$,
the equations above hold with $K_1( i^\prime j^\prime) = 0$. We
shall work in the closed subspace $H_B(ij) = H_S(i) = 0$ for all
$i$,$j$, where the recursion relation for $J^\prime(i^\prime
j^\prime)$ is a function
\begin{equation}\label{eq:64}
J^\prime(i^\prime j^\prime) = R(\{J(ij)\};K_1(i^\prime j^\prime))\,,
\end{equation}
with $\{J(ij)\} = \{J(i^\prime k_1),J(k_1 j^\prime),J(i^\prime
k_2),J(k_2 j^\prime)\}$ being the set of interaction constants in
the cluster, and $R$ given in Eqs.~\eqref{eq:10d} and
\eqref{eq:10e}.

If the interaction constants $J(ij)$ have a quenched probability
distribution ${\cal P}(J(ij))$, and the long-range bonds $K_q(ij)$
have a quenched probability distribution ${\cal Q}^{(q)}(K_q(ij))$,
the distribution ${\cal P}^{(n)}(J^\prime_{i^\prime j^\prime})$ for
the rescaled system after $n$ renormalization-group transformations
is given by the convolution
\begin{align}
&{\cal P}^{(n)}(J^\prime(i^\prime j^\prime)) =\nonumber\\
&\int \Bigl[\prod_{ij}^{i^\prime j^\prime} dJ(ij){\cal
P}^{(n-1)}(J(ij))\Bigr] dK_1(i^\prime j^\prime){\cal
Q}^{(n-1)}(K_1(i^\prime j^\prime))\nonumber\\
&\qquad \qquad \delta\left(J^\prime(i^\prime j^\prime) -
R(\{J_{ij}\};K_1(i^\prime j^\prime)\right)\,,\label{eq:65}
\end{align}
where the product runs over the nearest-neighbor bonds $ij$ in the
cluster between $i^\prime$ and $j^\prime$.  The long-range bond
distribution ${\cal Q}^{(n)}(K^\prime_1( i^\prime j^\prime))$ after
$n$ renormalization-group transformations is
\begin{align}
{\cal Q}^{(n)}(K^\prime_1(i^\prime j^\prime)) &= p\, \delta\left(K^\prime_1(i^\prime j^\prime)-J_0 (n+1)^{-\sigma}\right)\nonumber\\
&\qquad + (1-p) \delta\left(K^\prime_1(i^\prime
j^\prime)\right)\label{eq:66}\,.
\end{align}

The convolution in Eq.~\eqref{eq:65} is implemented numerically,
with the probability distribution ${\cal P}^{(n)}(J_{ij})$
represented by histograms, each histogram consisting of a bond
strength and its associated probability.  The initial distribution
${\cal P}^{(0)}(J_{ij})$ is a single histogram at $J_0$ with
probability $1$.  Since Eq.~\eqref{eq:65} is a convolution of five
probability distributions, computational storage limits can be used
most effectively by factoring it into an equivalent series of three
pairwise convolutions, each of which involves only two distributions
convoluted with an appropriate $R$ function.  Two types of pairwise
convolutions are required, a ``bond-moving'' convolution with
\begin{equation}\label{eq:67}
R_\text{bm}(J(i_1 j_1),J(i_2 j_2)) = J(i_1 j_1) + J(i_2 j_2)\,,
\end{equation}
and a decimation convolution with
\begin{equation}\label{eq:68}
R_\text{dc}(J(i_1 j_1),J(i_2 j_2)) =
\frac{1}{2}\ln\left(\frac{\cosh(J(i_1 j_1) + J(i_2
j_2))}{\cosh(J(i_1 j_1) - J(i_2 j_2))}\right)\,.
\end{equation}
Starting with the probability distribution ${\cal P}^{(n-1)}$, the
following series of pairwise convolutions gives the total
convolution of Eq.~\eqref{eq:65}:  (i) a decimation convolution of
${\cal P}^{(n-1)}$ with itself, yielding ${\cal P}_A$; (ii) a
bond-moving convolution of ${\cal P}_A$ with itself, yielding ${\cal
P}_B$; (iii) a bond-moving convolution of ${\cal P}_B$ with ${\cal
Q}^{(n-1)}$, yielding the final result ${\cal P}^{(n)}$.

Because the number of histograms representing the probability
distribution increases rapidly with each renormalization-group step,
we use a binning procedure~\cite{FalicovBerkerMcKay}:  before every
pairwise convolution, the histograms are placed on a grid, and all
histograms falling into the same grid cell are combined into a
single histogram in such a way that the average and the standard
deviation of the probability distribution are preserved. Histograms
falling outside the grid, representing a negligible part of the
total probability, are similarly combined into a single histogram.
Any histogram within a small neighborhood of a cell boundary is
proportionately shared between the adjacent cells.  After the
convolution, the original number of histograms is reattained. For
the results presented below we used 562,500 bins, requiring the
calculation of 562,500 local renormalization-group transformations
at every iteration.

For the thermodynamic densites $M_\alpha$, given by
\begin{equation}\label{eq:69}
M_\alpha = \frac{1}{N_\alpha} \sum_{ij} \frac{\partial \ln
Z}{\partial K_\alpha(ij)}\,,
\end{equation}
the chain rule yields conjugate recursion relations for the quenched
random system,
\begin{equation}\label{eq:70}
M_\alpha = b^{-d} \sum_\beta \frac{1}{N^\prime_\beta} \sum_{i^\prime
j^\prime} \frac{\partial \ln Z}{\partial K^\prime_\beta(i^\prime
j^\prime)}\, \sum^{i^\prime j^\prime}_{ij} \frac{N_\beta}{N_\alpha}
\frac{\partial K^\prime_\beta(i^\prime j^\prime)}{\partial
K_\alpha(ij)} \,,
\end{equation}
where the rightmost sum runs over nearest-neighbor bonds $ij$ in the
cluster between sites $i^\prime$ and $j^\prime$.  As an
approximation, this sum is replaced by its average value, so that
\begin{equation}\label{eq:72}
M_\alpha \approx b^{-d} \sum_\beta M_\beta^\prime
\overline{T}_{\beta\alpha}\quad \text{with}\quad
\overline{T}_{\beta\alpha} \equiv \sum^{i^\prime j^\prime}_{ij}
\frac{N_\beta}{N_\alpha} \overline{\frac{\partial
K^\prime_\beta(i^\prime j^\prime)}{\partial K_\alpha(ij)}}\,,
\end{equation}
Here the overbar denotes averaging over the probability
distributions of the interaction constants in the cluster shown in
Fig.~\ref{fig:cluster}.  Using the recursion relations in
Eq.~\eqref{eq:10d}, in the subspace ${H_B}_{ij} = H_S = 0$,
\begin{align}
\overline{\mathbf{T}} &= \overline{\begin{pmatrix}
4 & \sum^{i^\prime j^\prime}_{ij} {\frac{\partial G^\prime(i^\prime j^\prime)}{\partial J(ij)}} & 0 & 0\\
0 & \sum^{i^\prime j^\prime}_{ij} {\frac{\partial J^\prime(i^\prime j^\prime)}{\partial J(ij)}} & 0 & 0\\
0 & 0 & \sum^{i^\prime j^\prime}_{ij} {\frac{\partial
H^\prime_B(i^\prime j^\prime)}{\partial H_B(ij)}}
& \sum^{i^\prime j^\prime}_{i} {\frac{\partial H^\prime_B(i^\prime j^\prime)}{\partial H_{S}(i)}} \\
0 & 0 & 0 & \sum^{i^\prime}_{i} {\frac{\partial H^\prime_{S}(i^\prime)}{\partial H_{S}(i)}} \end{pmatrix}}\nonumber\\
&= \begin{pmatrix} 4 & 2\overline{u} & 0 & 0\\
0 & 2\overline{u} & 0 & 0 \\
0 & 0 & 2+2\overline{u} & {\frac{3\overline{w}}{2}}\\
0 & 0 & 0 & 1 \end{pmatrix}\,,\label{eq:73}
\end{align}
where
\begin{align}
u &= \frac{1}{2}(\tanh(J(i^\prime k_1) + J(k_1 j^\prime))\nonumber\\
&\qquad\qquad+ \tanh(J(i^\prime k_2) + J(k_2 j^\prime))),\nonumber\\
w &= \frac{\sinh(J(i^\prime k_1)+J(k_1 j^\prime)+J(i^\prime
k_2)+J(k_2j^\prime))} {2\cosh(J(i^\prime k_1)+J(k_1
j^\prime))\cosh(J(i^\prime k_2)+J(k_2 j^\prime))}\,.\label{eq:74}
\end{align}
For a fixed probability distribution of the renormalization-group
transformation (e.g., Fig.~\ref{fig:hist}), the thermal and magnetic
eigenvalues exponents $y_T$ and $y_H$ are obtained as
\begin{align}
b^{y_T} &= \sum^{i^\prime j^\prime}_{ij}
\overline{{\frac{\partial J^\prime(i^\prime j^\prime)}{\partial J(ij)}}} = 2\overline{u}\,,\nonumber\\
b^{y_H} &= \sum^{i^\prime j^\prime}_{ij} \overline{\frac{\partial
H^\prime_B(i^\prime j^\prime)}{\partial H_B(ij)}} =
2+2\overline{u}\,.\label{eq:75}
\end{align}

\subsubsection{Results}

\begin{figure}
\includegraphics[scale=1]{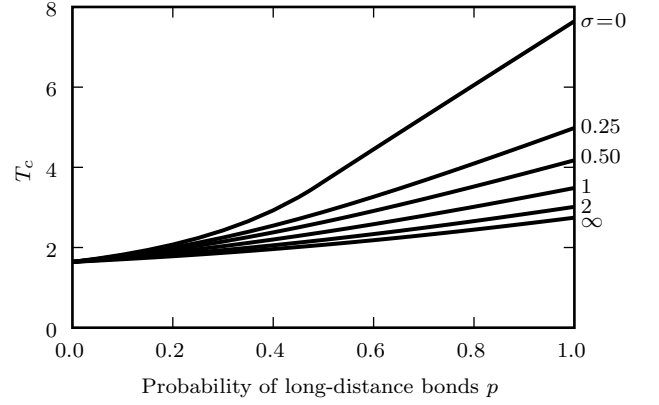}
\caption{Critical temperature $T_c$ as a function of the probability
of long-range bonds $p$, plotted for several values of the exponent
$\sigma$ characterizing the decay of the long-range bond
strengths.}\label{fig:temps}
\end{figure}

\begin{figure}
\includegraphics[scale=0.43]{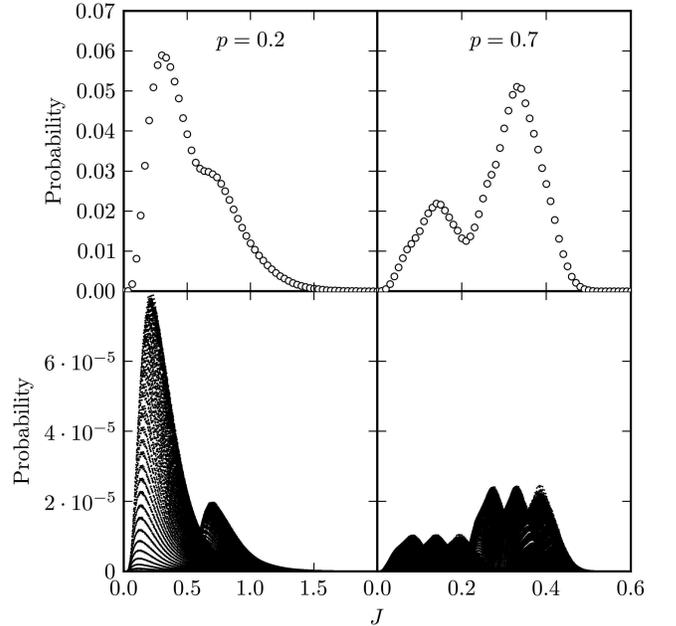}
\caption{Histograms of the unstable critical fixed probability
  distributions for $p=0.2$, $\sigma =0$ (left column) and $p=0.7$,
  $\sigma=0$ (right column).  The bottom panels show the actual
  histograms (numbering 1,128,002 in each case), while in the top
  panels the histograms are combined in order to clearly see the
  outlines of the probability distributions.}\label{fig:hist}
\end{figure}

\begin{figure}
\includegraphics[scale=1]{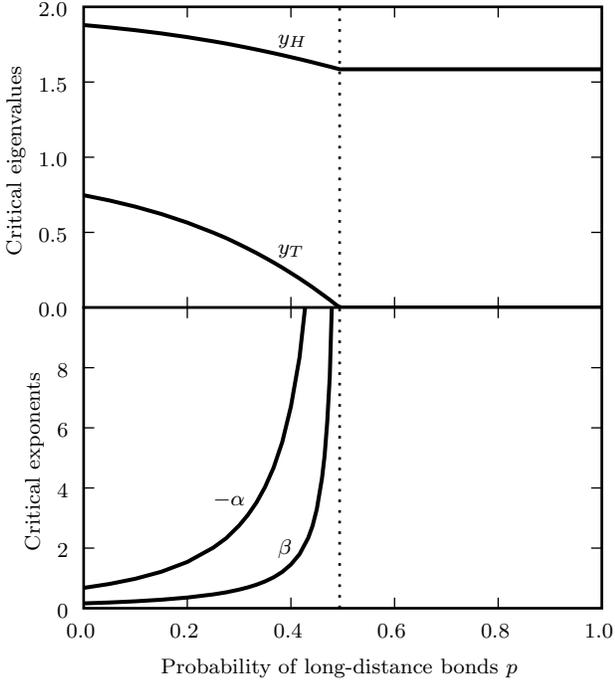}
\caption{The thermal and magnetic eigenvalues $y_T$ and $y_H$, and the
  corresponding specific heat and magnetic exponents $\alpha$ and
  $\beta$, as a function of $p$, for $\sigma =0$.  The probability $p =
  0.494$, marking the onset of exponentiated power-law critical
  behavior, is shown with a dotted line.}\label{fig:der}
\end{figure}

\begin{figure}
\includegraphics[scale=1]{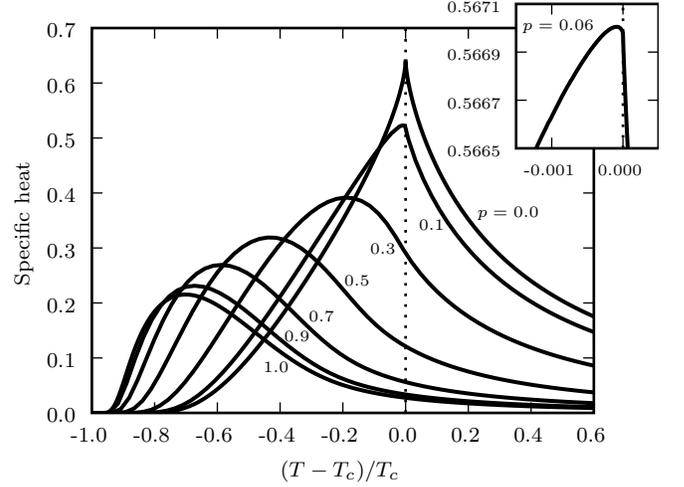}
\caption{Specific heat calculated for various probabilities $p$,
  plotted with respect to the normalized temperature $(T-T_c)/T_c$,
  with $\sigma =0$.  The vertical dotted line corresponds to the
  critical temperature $T=T_c$.  The inset shows a close-up of the
  specific heat near $T_c$ for $p=0.06$, showing both the
  infinite-slope singularity at $T=T_c$, and the analytic peak that
  appears for $T < T_c$ when $p \gtrsim 0.053$.}\label{fig:cdens}
\end{figure}

\begin{figure}
\includegraphics[scale=1]{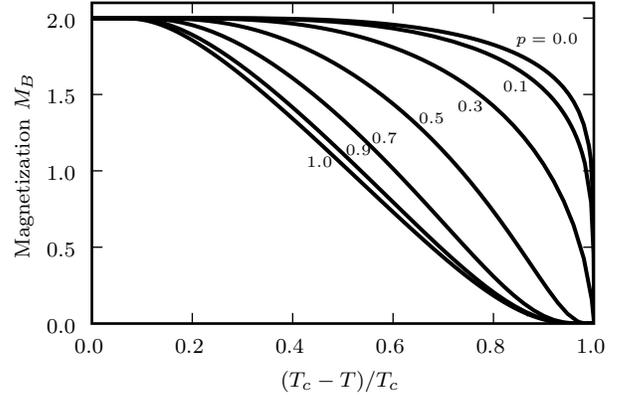}
\caption{Magnetization calculated for various probabilities $p$,
plotted with respect to the normalized temperature variable
$(T_c-T)/T_c$, with $\sigma =0$.  For $p > 0.363$, note the unusual
flat onset at $T_c$.}\label{fig:mdens}
\end{figure}

The quenched random system critical temperatures $T_c(p) =
J^{-1}_c(p)$ are shown as a function of $p$, in
Fig.~\ref{fig:temps}, for several values of the decay exponent
$\sigma$.  For any $\sigma > 0$, in renormalization-group
trajectories starting near $T_c$, the probability distribution
${\cal P}(J(ij))$ spends many iterations in the vicinity of the
unstable critical fixed distribution which is a delta function at
$J_c(p=0)$, the $p=0$ critical interaction strength given by
Eq.~\eqref{eq:11}. Similarly to the results of the $p=1$ case given
above, when $\sigma > 0$, the critical behavior for all $p$ is that
of $p=0$, though with a rapidly decreasing critical region as $p \to
1$ and $\sigma \to 0$.

On the other hand, for $\sigma = 0$, a variety of critical behaviors
occurs as $p$ ranges from 0 to 1.  The unstable critical fixed
distribution has a non-trivial structure which depends on $p$, two
examples of which are shown in Fig.~\ref{fig:hist}.  The eigenvalue
exponents $y_T$ and $y_H$ from the critical fixed distributions
change continuously with $p$ (Fig.~\ref{fig:der}), and with them the
critical exponents characterizing the phase transition.  As $p$ is
increased from zero, both $y_T$ and $y_H$ decrease from their $p=0$
values, attaining their $p=1$ values of $y_T = 0$ and $y_H = \ln
3/\ln 2$ at $p = 0.494$. Thus, the system has two distinct regimes
of criticalities.  For $p < 0.494$ the critical behavior is
described by power laws with exponents $\nu = 1/y_T$, $\alpha =
(2y_T-d)/y_T$, $\beta = (d-y_H)/y_T$, and $\gamma = (2y_H-d)/y_T$.
As $y_T \to 0$ with $p \to 0.494$, the exponents blow up as $\nu \to
\infty$, $\alpha \to -\infty$, $\beta \to \infty$, and $\gamma \to
\infty$. For $p \geq 0.494$ the critical behavior is that of the
$p=1, \sigma = 0$ case given above, with exponentiated power laws of
the thermodynamic quantities, and the high-temperature phase has
infinite correlation length.  The onset of exponentiated power-law
critical behavior at $p = 0.494$, due to the influence of the
long-range bonds, in fact corresponds to a change in the geometrical
features of the network. As we have noted in Fig.~\ref{fig4}, for $p
\gtrsim 0.5$ the average path length $\ell$ has a small world
character, $\ell \sim \ln N_n$, while for smaller $p$ it increases
more rapidly like $N^{1/d}_n$, as in a regular lattice.

The spectrum of critical behaviors for varying $p$ at $\sigma = 0$
is illustrated in Figs.~\ref{fig:cdens} and \ref{fig:mdens} for the
specific heat and magnetization versus temperature for different
values of $p$. With increasing $p$ from $p=0$, the low-temperature
analytic peak, due to the saturation of long-range order, as
mentioned above, appears at $p \approx 0.053$, as the
low-temperature amplitude of the critical cusp changes sign, and
shifts to lower temperatures as $p$ further increases. At the
critical-point singularity, with increasing $p$ from $p=0$, the
specific heat exponent $\alpha$ continuously decreases from its
$p=0$ value of $-0.677$: The cusp disappears at $p=0.105$ as
$\alpha$ crosses $-1$, so that the specific heat acquires a
continuous slope at criticality, but all higher derivatives remain
divergent.  The second derivative at criticality also becomes
continuous, all higher derivatives remaining divergent, at $p=0.249$
as $\alpha$ crosses $-2$.  Thus, as $\alpha$ crosses the consecutive
negative integers at at $p=0.105$, $0.249$, $0.312$, $0.349,\ldots$,
the divergence begins at a higher derivative, until the accumulation
point at $p=0.494$, where $\alpha$ reaches $-\infty$, and the
essential singularity occurs for the higher values of $p$.

In the magnetization, with increasing $p$ from $p=0$, the critical
exponent $\beta$ continuously increases from its $p=0$ value of
$0.162$.  Thus, the slope at criticality changes from infinity to zero
at $p=0.363$ as $\beta$ crosses $1$, but all higher derivatives of the
magnetization remain divergent. The second derivative at criticality
also becomes zero, all higher derivatives remaining divergent, at
$p=0.424$ as $\beta$ crosses $2$.  At each crossing of a positive
integer by $\beta$, at $p=0.363$, $0.424$, $0.446$, $0.457,\ldots$,
the zeros extend to one higher derivative and the divergence begins at
one higher derivative, until the accumulation point at $p=0.494$,
where $\beta$ reaches $\infty$, and the essential singularity occurs
for the higher values of $p$.

\section{Conclusions}

In summary, we have introduced a scale-free
hierarchical-lattice network model exhibiting a range of geometric
and thermodynamic properties as we vary the probability $p$ of the
long-distance bonds.  When $p=0$, our network is unclustered and the
average shortest-path length $\ell$ scales like a power-law in the
number of sites, $\ell \sim N^{1/2}$, resembling in these respects a
regular lattice.  This resemblance also holds for the critical
behavior of the ferromagnetic Ising model on the $p=0$ network,
which shows typical power-law singularities in specific heat,
magnetization, and susceptibility.  For small concentrations of
long-distance bonds, $p < 0.494$, this picture does not change
radically: the clustering coefficient increases nearly linearly with
$p$, the average shortest-path length continues to have power-law
scaling with lattice size, and the critical exponents of the Ising
model vary continuously with $p$.  As we approach $p=0.494$,
however, the magnitudes of these exponents blow up, and we have an
unexpected crossover to a completely different regime of critical
behavior for $p \ge 0.494$.  We find a highly unusual infinite-order
phase transition, an inverted Berezinskii-Kosterlitz-Thouless
singularity between a low-temperature phase with nonzero
magnetization and finite correlation length, and a high-temperature
phase with zero magnetization but infinite correlation length and
power-law decay of correlations throughout the phase.  This slow
decay of correlations in the disordered phase is a direct
consequence of the underlying lattice topology, since large enough
concentrations of long-distance bonds significantly reduce the
shortest-path length between any pair of sites.  Indeed for $p \ge
0.494$ the network shows the small-world effect, with the average
shortest-path scaling logarithmically as $\ell \sim \ln N$.

In determining these geometric features and critical behaviors of
our model, we were aided by the deterministic, hierarchical nature
of the network construction.  This allowed us to derive analytic
expressions for many of the network characteristics: the degree
distribution and the clustering coefficient at all $p$, and the
average shortest-path length for $p=0$ and $p=1$.  In addition, we
were able to formulate exact renormalization-group transformations
for the Ising model on the network, even with a quenched random
distribution of the long-distance bonds.  The present model was
designed to incorporate just some of the distinctive properties of
real-world networks: scale-free degree distribution, high clustering
coefficient, and small-world effect.  But such hierarchical-lattice
models could also be tailored to capture other empirical properties,
like the modular, community structure of networks.  Using techniques
similar to the ones we applied to the Ising model, one could develop
exact renormalization-group approaches to other interesting
statistical physics systems, for example percolation models relevant
to epidemic spreading.  Finally, non-equilibrium study~\cite{Candia}
of our model should yield interesting new results.

\begin{acknowledgments}

This research was supported by the Scientific and Technical Research
Council of Turkey (T\"UBITAK) and by the Academy of Sciences of
Turkey.

\end{acknowledgments}

\appendix

\section{Derivation of network characteristics}

\subsection{Average clustering coefficient $C_m$}

\begin{figure}[t]
\centering\includegraphics*[scale=0.42]{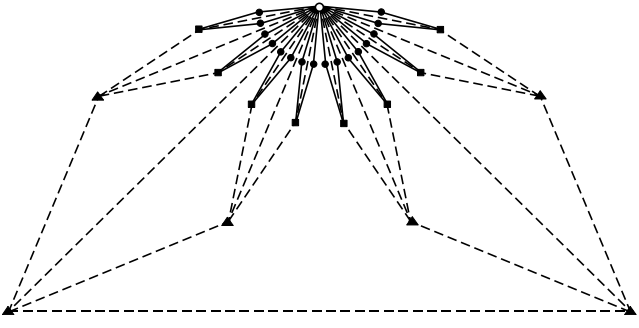}
\caption{The open circle on top represents a site with
$k_\text{nn}=2^m$, $k_\text{ld} = 2^m -2$, for $m=4$, and the figure
shows all possible neighbors of this site, together with the
possible bonds among those neighbors.  The $2^m$ sites connected to
the top site by nearest-neighbor bonds are drawn as filled circles,
the $2^{m-1}$ sites potentially connected to the top site by the
shortest long-range bonds are drawn as squares, and the other
$2^{m-1}-2$ potential neighbors as triangles.}\label{apfig1}
\end{figure}

Consider a site in the infinite lattice with $k_\text{nn}=2^m$,
$k_\text{ld} = 2^m -2$, and $m>1$, its possibly connected sites, and
all possible bonds among those sites.  The $m=4$ case is shown in
Fig.~\ref{apfig1}.  To calculate the average clustering coefficient
$C_m$ of such a site, we must consider the various configurations of
long-range bonds among the possibly connected sites. The $2^m-2$
potential long-range bonds emanating from the original site we
divide into two categories:  the $2^{m-1}$ ``shortest'' ones (to the
sites marked as squares in Fig.~\ref{apfig1}), and the remaining
$2^{m-1}-2$ bonds (to the sites marked as triangles in
Fig.~\ref{apfig1}).  The probability for $r$ bonds of the first
category and $r^\prime$ bonds of the second category is
\begin{equation}\label{eq:app1}
P_{r,r^\prime}=\binom{2^{m-1}}{r} \binom{2^{m-1}-2}{r^\prime}  p^{r+r^\prime}(1-p)^{2^m-2-r-r^\prime}.
\end{equation}
For a given $r$ and $r^\prime$, the site has
$k_{r,r^\prime}=2^m+r+r^\prime$ connected sites, so its average
clustering coefficient is
\begin{equation}\label{eq:app2}
C_m = \sum_{r=0}^{2^{m-1}}\sum_{r^\prime=0}^{2^{m-1}-2} P_{r,r^\prime} \frac{B_{r,r^\prime}}{k_{r,r^\prime}(k_{r,r^\prime}-1)/2}\,,
\end{equation}
where $B_{r,r^\prime}$ is the average number of bonds which actually
exist among the $k_{r,r^\prime}$ sites connected to the original
site.  Each of the $r$ bonds of the first category contributes two
to $B_{r,r^\prime}$, as can be seen in Fig.~\ref{apfig1}, where
there are nearest-neighbor bonds connecting every square site to two
of the $2^m$ filled circle sites.  There are $\binom{r+r^\prime}{2}$
ways of choosing pairs among the $r+r^\prime$ neighbors connected to
the main site by long-range bonds, but of these pairs, only a
fraction $(2^m-3)/\binom{2^m-2}{2}$ corresponds to possible
long-range bonds between those neighbors, and of these possible
bonds on average only a fraction $p$ will actually exist.  So the
total expression for $B_{r,r^\prime}$ is
\begin{equation}\label{eq:app3}
B_{r,r^\prime} = 2r + p\binom{r+r^\prime}{2} \frac{2^m-3}{\binom{2^m-2}{2}}\,.
\end{equation}
Putting together Eqs.~\eqref{eq:app1}-\eqref{eq:app3} yields the
expression for $C_m$ in Eq.~\eqref{eq:3}.

\vfill
\subsection{Average shortest-path length $\ell_n$\\ for $p=0$ and $p=1$}

\begin{figure}[t]
  \centering\includegraphics*[scale=0.35]{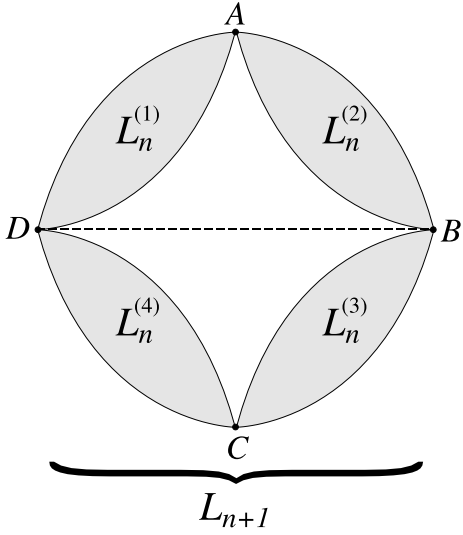}
  \caption{For $p=0$ or $p=1$, the lattice after $n+1$
    construction steps, $L_{n+1}$, is composed of four copies of $L_n$
    connected to one another as above.  The $p=1$ case is shown
    above; for $p=0$ the horizontal long-range bond is absent.}\label{apfig2}
\end{figure}

Let us denote the set of sites making up the lattice after $n$
construction steps as $L_n$.  Then the average shortest-path length
for $L_n$ is defined to be:
\begin{equation}\label{eq:app4}
  \ell_n  = \frac{S_n}{N_n(N_n-1)/2}\,,
\end{equation}
where
\begin{equation}\label{eq:app5}
  S_n = \sum_{i,j \in L_n} d_{ij}\,,
\end{equation}
and $d_{ij}$ is the length of the shortest path between sites $i$ and
$j$.  For the cases $p=0$ and $p=1$, the lattice has a self-similar
structure that allows one to calculate $\ell_n$ analytically.  As
shown in Fig.~\ref{apfig2}, the lattice $L_{n+1}$ in these cases is
composed of four copies of $L_n$ connected at the edges, which we
label $L_n^{(\alpha)}$, $\alpha=1,\ldots,4$.  We can write the sum over all
shortest paths $S_{n+1}$ as
\begin{equation}\label{eq:app6}
  S_{n+1} = 4S_n + \Delta_n\,,
\end{equation}
where $\Delta_n$ is the sum over all shortest paths whose endpoints
are not in the same $L_n$ branch. The solution of
Eq.~\eqref{eq:app6} is
\begin{equation}\label{eq:app8}
  S_n = 4^{n-1} S_1 + \sum_{m=1}^{n-1} 4^{n-m-1} \Delta_m\,.
\end{equation}
The paths that contribute to $\Delta_n$ must all go through at least
one of the four edge sites ($A$, $B$, $C$, $D$) at which the
different $L_n$ branches are connected.  The analytical expression
for $\Delta_n$, which we call the crossing paths, are found below
for $p=0$ and $p=1$.

\subsubsection{Crossing paths $\Delta_n$ for $p=0$}

Let $\Delta_n^{\alpha,\beta}$ denote the sum of all shortest paths
with endpoints in $L_n^{(\alpha)}$ and $L_n^{(\beta)}$. If
$L_n^{(\alpha)}$ and $L_n^{(\beta)}$ meet at an edge site,
$\Delta_n^{\alpha,\beta}$ excludes paths where either endpoint is
that shared edge site.  If $L_n^{(\alpha)}$ and $L_n^{(\beta)}$ do
not meet, $\Delta_n^{\alpha,\beta}$ excludes paths where either
endpoint is any edge site.  Then the total sum $\Delta_n$ is given
by
\begin{align}
\Delta_n =& \,\Delta_n^{1,2} + \Delta_n^{2,3} + \Delta_n^{3,4}
+\Delta_n^{4,1}+ \Delta_n^{1,3} + \Delta_n^{2,4}\nonumber\\
& - 2\cdot 2^{n+1}\,.
\label{eq:app10}
\end{align}
The last term at the end compensates for the overcounting of certain
paths: the shortest path between $A$ and $C$, with length $2^{n+1}$,
is included in both $\Delta_n^{2,3}$ and $\Delta_n^{4,1}$. Similarly
the shortest path between $B$ and $D$, also with length $2^{n+1}$,
is included in both $\Delta_n^{1,2}$ and $\Delta_n^{3,4}$.

By symmetry, $\Delta_n^{1,2} = \Delta_n^{2,3} = \Delta_n^{3,4} =
\Delta_n^{4,1}$ and $\Delta_n^{1,3} = \Delta_n^{2,4}$, so that
\begin{equation}\label{eq:app11}
\Delta_n = 4 \Delta_n^{1,2} + 2\Delta_n^{1,3} - 2\cdot 2^{n+1}\,.
\end{equation}
$\Delta_n^{1,2}$ is given by the sum
\begin{align}
  \Delta_n^{1,2} &= \sum_{\substack{i \in L_n^{(1)},\,\,j\in
      L_n^{(2)}\\ i,j \ne A}} d_{ij}\nonumber\\
  &= \sum_{\substack{i \in L_n^{(1)},\,\,j\in L_n^{(2)}\\i,j \ne A}} (d_{iA} + d_{Aj}) \nonumber\\
  &= (N_n-1)\sum_{i \in L_n^{(1)}} d_{iA} + (N_n-1) \sum_{j
    \in L_n^{(2)}} d_{Aj} \nonumber\\
  &= 2(N_n-1)\sum_{i \in L_n^{(1)}} d_{iA}\,,
\label{eq:app12}
\end{align}
where we have used $\sum_{i \in L_n^{(1)}} d_{iA} = \sum_{j \in
L_n^{(2)}} d_{Aj}$.  To find $\sum_{i \in
  L_n^{(1)}} d_{iA}$, we examine the structure of the
hierarchical lattice at the $n$th level.  $L_n^{(1)}$ contains
$\nu_n(m)$ points with $d_{iA} = m$, where $1 \le m \le 2^n$, and
$\nu_n(m)$ can be written recursively as follows:
\begin{equation}\label{eq:app13}
\nu_n(m) = \begin{cases} 2^n & \text{if $m$ is odd}\,,\\
\nu_{n-1}(m/2) & \text{if $m$ is even}\,.\end{cases}
\end{equation}
Expressing $\sum_{i \in
  L_n^{(1)}} d_{iA}$ in terms of $\nu_n(m)$,
\begin{equation}\label{eq:app14}
f_n \equiv \sum_{i \in L_n^{(1)}} d_{iA} = \sum_{m=1}^{2^n} m
\nu_n(m)\,.
\end{equation}
Eqs.~\eqref{eq:app13} and \eqref{eq:app14} relate $f_n$ and
$f_{n-1}$, allowing the solution of $f_n$ by induction:
\begin{align}
f_n &= \sum_{k=1}^{2^{n-1}} (2k-1) 2^n + \sum_{k=1}^{2^{n-1}} 2k
\nu_{n-1}(k)\nonumber\\
&= 2^{3n-2} + 2f_{n-1}\nonumber\\
&= \sum_{k=0}^{n-2} 2^k 2^{3(n-k)-2} + 2^{n-1}f_1 \nonumber\\
&= \frac{1}{3}2^n (2+4^n)\,, \label{eq:app15}
\end{align}
where we have used $f_1 = \nu_1(1) + 2\nu_1(2)= 4$.  Substituting
Eq.~\eqref{eq:app15} and $N_n = \frac{2}{3}(2+4^n)$ into
Eq.~\eqref{eq:app12},
\begin{equation}\label{eq:app16}
\Delta_n^{1,2} = \frac{1}{9}2^{1+n}(1+2^{1+2n})(2+ 4^n)\,.
\end{equation}

Proceeding similarly,
\begin{align}
\Delta_n^{1,3} =& \sum_{\substack{i \in L_n^{(1)},\,\,j\in
      L_n^{(3)}\\ i\ne A, D,\,\,j\ne B, C}} d_{ij}\nonumber\\
=& \sum_{\substack{i \in L_n^{(1)},\,\,j\in
      L_n^{(3)}\\ i\ne A,\,\,j\ne B,\,\,d_{iA}+d_{jB} < 2^n}} (d_{iA}
  + 2^n + d_{jB})\nonumber\\
&+\sum_{\substack{i \in L_n^{(1)},\,\,j\in
      L_n^{(3)}\\ i\ne D,\,\,j\ne C,\,\,d_{iD}+d_{jC} < 2^n}} (d_{iD}
  + 2^n + d_{jC})\nonumber\\
&+\sum_{\substack{i \in L_n^{(1)},\,\,j\in
      L_n^{(3)}\\ i\ne A,\,\,j\ne B,\,\,d_{iA}+d_{jB} = 2^n}}
  2^{n+1}\,.
\label{eq:app17}
\end{align}
The first and second terms are equal and denoted by $g_n$, and the
third term is denoted by $h_n$, so that $\Delta_n^{1,3} = 2g_n +
h_n$. The quantity $g_n$ is evaluated as follows:
\begin{align}
  g_n =& \sum_{m=1}^{2^n-2}\:\sum_{m^\prime = 1}^{2^n-1-m} \nu_n(m)\nu_n(m^\prime)(m+2^n+m^\prime)\nonumber\\
  =& \sum_{k=1}^{2^{n-1}-2}\:\sum_{k^\prime = 1}^{2^{n-1}-1-k} \nu_{n-1}(k)\nu_{n-1}(k^\prime)(2k+2^n+2k^\prime)\nonumber\\
  &+ \sum_{k=1}^{2^{n-1}-1}\:\sum_{k^\prime = 1}^{2^{n-1}-k} \nu_{n-1}(k)2^n (2k+2^n+2k^\prime-1)\nonumber\\
  &+ \sum_{k=1}^{2^{n-1}-1}\:\sum_{k^\prime = 1}^{2^{n-1}-k} 2^n \nu_{n-1}(k^\prime) (2k-1+2^n+2k^\prime)\nonumber\\
  &+ \sum_{k=1}^{2^{n-1}-1}\:\sum_{k^\prime = 1}^{2^{n-1}-k} 2^{2n}
  (2k-1+2^n+2k^\prime-1)\,.
\label{eq:app18}
\end{align}
The fourth term can be summed directly, yielding
\begin{align}
8^{n-1}(2^n-2)(5\cdot 2^n-2)/3\,. \label{eq:app19}
\end{align}
The second and third terms in Eq.~\eqref{eq:app18} are equal and can
be simplified by first summing over $k^\prime$, yielding
\begin{align}
&2^{n-2}\sum_{k=1}^{2^{n-1}-1} \nu_{n-1}(k) (3\cdot 4^n - 2^{n+2} k
- 4k^2)\,. \label{eq:app20}
\end{align}
For use in Eq.~\eqref{eq:app20}, $\sum_{k=1}^{2^{n-1}-1}
\nu_{n-1}(k) = N_{n-1}-2$, and using Eq.~\eqref{eq:app15},
\begin{align}
  \sum_{k=1}^{2^{n-1}-1} k \nu_{n-1}(k) &=
  \sum_{k=1}^{2^{n-1}} k \nu_{n-1}(k) - 2^{n-1}\nonumber\\
  &= 2^{n-1}(4^{n-1}-1)/3\,.
 \label{eq:app21}
\end{align}
Analogously to Eq.~\eqref{eq:app15}, we find
\begin{align}
&\sum_{k=1}^{2^{n-1}-1} k^2 \nu_{n-1}(k)=
\frac{1}{9}2^{2n-3}(14+4^n-3n-3) - 2^n\,. \label{eq:app22}
\end{align}
With the latter results, Eq.~\eqref{eq:app20} becomes
\begin{align}
8^{n-1}(-23+5\cdot 4^n+3n)/9\,. \label{eq:app23}
\end{align}
With Eqs.~\eqref{eq:app19} and \eqref{eq:app23},
Eq.~\eqref{eq:app18} becomes
\begin{align}
  g_n = 2g_{n-1} + 8^{n-1}(-34-9\cdot 2^{n+2}+25\cdot 4^n+6n)/9\,.
\label{eq:app24}
\end{align}
Using $g_1 = 0$, Eq.~\eqref{eq:app24} is solved inductively:
\begin{align}
g_n &= 2^n (164-126\cdot 4^n -108\cdot 8^n+
  70\cdot 16^n\nonumber\\
&\qquad +21\cdot 4^n n)/189\,. \label{eq:app25}
\end{align}
All that is left to find an expression for $\Delta_n^{1,3}$ is
to evaluate
\begin{align}
h_n &= 2^{n+1} \sum_{m=1}^{2^n-1} \nu_n(m) \nu_n(2^n-m)\nonumber\\
&= 2^{n+1} \sum_{m=1}^{2^n-1} \nu^2_n(m)\nonumber\\
&= 2^{n+1} \left[\sum_{k=1}^{2^{n-1}} 4^n  + \sum_{k=1}^{2^{n-1}-1}
  \nu^2_{n-1}(k)\right]\nonumber\\
&= 16^n + 2h_{n-1}\,,
\label{eq:app26}
\end{align}
where the symmetry $\nu_n(m) = \nu_n(2^n-m)$ was used.  Using $h_1 =
16$, Eq.~\eqref{eq:app26} is solved inductively:
\begin{equation}
h_n = 2^{n+3}(8^n-1)/7\,. \label{eq:app27}
\end{equation}
From Eqs.~\eqref{eq:app25} and \eqref{eq:app27},
\begin{align}
\Delta_n^{1,3} &= 2^{n+1}(8-18\cdot 4^n
  +10\cdot 16^n +3\cdot 4^n n)/27\,.
\label{eq:app28}
\end{align}
Substituting Eqs.~\eqref{eq:app16} and \eqref{eq:app28} into
Eq.~\eqref{eq:app11}, we obtain the final expression for the
crossing paths $\Delta_n$ when $p=0$:
\begin{equation}
\Delta_n= 2^{2+n}(-7+3(4+n)4^{n}+22\cdot 16^n)/27\,.
\label{eq:app29}
\end{equation}
\vspace{1em}

\subsubsection{Crossing paths $\Delta_n$ for $p=1$}

In the $p=1$ case, $\Delta_n$ is
\begin{align}
\Delta_n^{1,2} + \Delta_n^{2,3} + \Delta_n^{3,4} +\Delta_n^{4,1}+
\Delta_n^{1,3} + \Delta_n^{2,4}- 3\,. \label{eq:app30}
\end{align}
The last term compensates for the overcounting of the shortest path
between $A$ and $C$, with length $2$, and the shortest path between
$B$ and $D$, with length $1$.

Again by symmetry, $\Delta_n^{1,2} = \Delta_n^{3,4}$,
$\Delta_n^{2,3} = \Delta_n^{4,1}$, and $\Delta_n^{1,3} =
\Delta_n^{2,4}$, so that
\begin{equation}\label{eq:app31}
\Delta_n = 2 \Delta_n^{1,2} + 2\Delta_n^{2,3} + 2\Delta_n^{1,3} - 3\,.
\end{equation}
We define
\begin{align}
d_n^\text{tot} &\equiv \sum_{i \in L_n^{(1)}} d_{iA}\,,\nonumber\\
d_n^\text{near} &\equiv \sum_{\substack{i \in L_n^{(1)}\\ d_{iA} <
    d_{iD}}} d_{iA}\,,\qquad N_n^\text{near} \equiv \sum_{\substack{i
    \in L_n^{(1)}\\ d_{iA} < d_{iD}}} 1\,,\nonumber\\
d_n^\text{mid} &\equiv \sum_{\substack{i \in L_n^{(1)}\\ d_{iA} =
    d_{iD}}} d_{iA}\,,\qquad N_n^\text{mid} \equiv \sum_{\substack{i
    \in L_n^{(1)}\\ d_{iA} = d_{iD}}} 1\,,\nonumber\\
d_n^\text{far} &\equiv \sum_{\substack{i \in L_n^{(1)}\\ d_{iA} >
    d_{iD}}} d_{iA}\,,\qquad N_n^\text{far} \equiv \sum_{\substack{i
    \in L_n^{(1)}\\ d_{iA} > d_{iD}}} 1\,,
\label{eq:app32}
\end{align}
so that $d_n^\text{tot} = d_n^\text{near} + d_n^\text{mid} +
d_n^\text{far}$ and $N_n = N_n^\text{near} + N_n^\text{mid} +
N_n^\text{far}$.  By symmetry $N_n^\text{near} = N_n^\text{far}$.
Thus, {\footnotesize
\begin{align}
\Delta_n^{1,2} &= \sum_{\substack{i \in L_n^{(1)},\,\,j\in
      L_n^{(2)}\\ i,j \ne A}} d_{ij}
= \sum_{\substack{i \in L_n^{(1)},\,\,j\in
      L_n^{(2)}\\ i,j \ne A,\,\, d_{iA} \le d_{iD}}}
  (d_{iA}+d_{Aj})\nonumber\\
&\quad+ \sum_{\substack{i \in L_n^{(1)},\,\,j\in
      L_n^{(2)},\,\,i,j\ne A \\d_{iA} > d_{iD},\,\,d_{jA}\le d_{jB}}}
  (d_{iA}+d_{Aj})\nonumber\\
&\quad+ \sum_{\substack{i \in L_n^{(1)},\,\,j\in
      L_n^{(2)},\,\,i,j\ne A \\d_{iA} > d_{iD},\,\,d_{jA}> d_{jB}}}
  (d_{iD}+1+d_{Bj})\nonumber\\
 &= \sum_{\substack{i \in
      L_n^{(1)},\,\, i \ne A\\ d_{iA} \le d_{iD}}}
  \left[(N_n-1)d_{iA}+d_n^\text{tot}\right]\nonumber\\
&\quad+\sum_{\substack{i \in
      L_n^{(1)},\,\, i \ne A\\ d_{iA} > d_{iD}}}
  \left[(N_n^\text{near}+N_n^\text{mid}-1)d_{iA}+d_n^\text{near}+d_n^\text{mid}\right]\nonumber\\
&\quad+\sum_{\substack{i \in L_n^{(1)},\,\,i\ne A \\d_{iA} > d_{iD}}}
\left[N_n^\text{near} (d_{iD} + 1) + d_n^\text{near}\right]\nonumber\\
&= (N_n-1)(d_n^\text{near}+d_n^\text{mid})
+(N_n^\text{near}+N_n^\text{mid}-1)d_n^\text{tot}\nonumber\\
&\quad+(N_n^\text{near}+N_n^\text{mid}-1)d_n^\text{far}
+N_n^\text{near}(d_n^\text{near}+d_n^\text{mid})\nonumber\\
&\quad+N_n^\text{near} (d_n^\text{near} + N_n^\text{near}) +
N_n^\text{near} d_n^\text{near}\,.
\label{eq:app33}
\end{align}}

The horizontal long-range bond does not affect $\Delta_n^{2,3}$, so
that Eq.~\eqref{eq:app12} still holds, $\Delta_n^{2,3} =
2(N_n-1)d_n^\text{tot}$.  Finally,
{\footnotesize
\begin{align}
\Delta_n^{1,3} &= \sum_{\substack{i \in L_n^{(1)},\,\,j\in
      L_n^{(3)}\\ i\ne A,D,\,\,j\ne B,C}} d_{ij}
= \sum_{\substack{i \in L_n^{(1)},\,\,j\in
      L_n^{(3)}\\ i\ne A,D,\,\,j\ne B,C,\,\, d_{jB} > d_{jC}}}
  (d_{iD}+1 + d_{Cj})\nonumber\\
&\quad+\sum_{\substack{i \in L_n^{(1)},\,\,j\in
      L_n^{(3)},\,\,i\ne A,D\\ j\ne B,C,\,\, d_{iA} \ge
      d_{iD},\,\,d_{jB} \le d_{jC}}}  (d_{iD} +1 + d_{Bj})\nonumber\\
&\quad+\sum_{\substack{i \in L_n^{(1)},\,\,j\in
      L_n^{(3)},\,\,i\ne A,D\\ j\ne B,C,\,\, d_{iA} <
      d_{iD},\,\,d_{jB} \le d_{jC}}}  (d_{iA} +1 + d_{Bj})\nonumber\\
&= \sum_{i \in L_n^{(1)},\,\,i\ne A,D}
  \left[(N_n^\text{near}-1)(d_{iD}+1) + d_n^\text{near}\right]\nonumber\\
&\quad+\sum_{\substack{i \in L_n^{(1)},\,\,i\ne A,D\\ d_{iA} \ge
      d_{iD}}} \left[(N_n^\text{near}+N_n^\text{mid}-1)(d_{iD}
    +1)\right.\nonumber\\
&\quad\qquad\qquad \left.+ d_n^\text{near}+d_n^\text{mid}\right]\nonumber\\
&\quad+\sum_{\substack{i \in L_n^{(1)},\,\,i\ne A,D\\ d_{iA} <
      d_{iD}}}  \left[(N_n^\text{near}+N_n^\text{mid}-1)(d_{iA}
    +1)\right.\nonumber\\
&\quad\qquad\qquad
\left.+d_n^\text{near}+d_n^\text{mid}\right]\nonumber\\
&=(N_n^\text{near}-1)(d_n^\text{tot}-1+N_n-2) +
(N_n-2)d_n^\text{near}\nonumber\\
&\quad +(N_n^\text{near}+N_n^\text{mid}-1)(d_n^\text{near}
 +d_n^\text{mid} + N_n^\text{near}+N_n^\text{mid}-1)\nonumber\\
&\quad\qquad\qquad +
(N_n^\text{near}+N_n^\text{mid}-1)(d_n^\text{near}+d_n^\text{mid})\nonumber\\
&\quad + (N_n^\text{near}+N_n^\text{mid}-1)(d_n^\text{near}
    +N_n^\text{near}-1)\nonumber\\
&\quad\qquad\qquad +(N_n^\text{near}-1)(d_n^\text{near}+d_n^\text{mid})\,.
\label{eq:app35}
\end{align}}Having $\Delta_n^{1,2}$, $\Delta_n^{2,3}$, and
$\Delta_n^{1,3}$ in terms of the quantities in Eq.~\eqref{eq:app32},
the next step is to explicitly determine these quantities.

\begin{figure}[t]
  \centering\includegraphics*[scale=1.7]{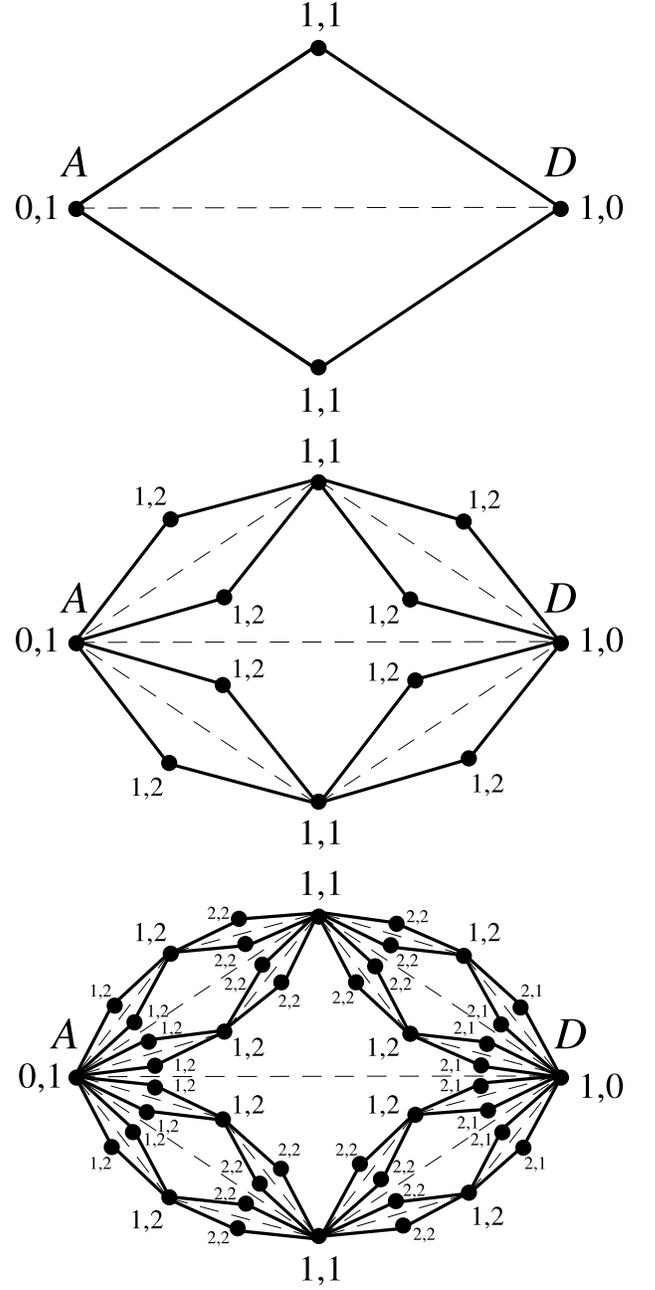}
  \caption{The first three construction steps of the lattice with $p=1$,
 with the sites labeled by ordered pairs denoting the
 shortest-path distance to the left and right edge sites.}\label{apfig3}
\end{figure}

We consider a site $i \in L_n^{(1)}$ and the shortest-path distances
to the edges, $d_{iA}$ and $d_{iD}$.  If the site was added to the
lattice at the $n$th construction step, the values of $d_{iA}$ and
$d_{iD}$ do not change at subsequent steps, since the shortest path
to the edge sites is always along the bonds added earliest.  We see
this in Fig.~\ref{apfig3}, where the sites are labeled by the
ordered pairs $d_{iD},\,d_{iA}$, for the first three construction
steps.  We denote by $a^n_{m,m^\prime}$ the number of sites added at
the $n$th construction step which have $d_{iA} = m$, $d_{iD} =
m^\prime$.  Since $A$ and $D$ are connected by a long-range bond,
$m^\prime$ and $m$ can differ by at most 1.  Thus for a given $m$
there are three categories of sites added at step $n$, respectively
numbering $a^n_{m,m+1}, a^n_{m,m}$, and $a^n_{m+1,m}$. By symmetry
$a^n_{m+1,m} = a^n_{m,m+1}$. The $m$, $m^\prime$ values of sites
added at step $n$ depend on the neighboring sites, which were added
at previous construction steps. For example, there are $2^n$ sites
added at the $n$th step ($n \ge 2$) which are nearest-neighbors of
site $A$, so these new sites have $m=1$, $m^\prime =2$, giving
$a^n_{1,2} = 2^n$. Sites with $m=1$, $m^\prime=2$ will in turn get
neighbors with $m=2$, $m^\prime=3$ in subsequent steps.  The
relationship between $a^n_{2,3}$ and $a^k_{1,2}$ for $k < n$ is
\begin{equation}\label{eq:app36}
a^n_{2,3} = \sum_{k=2}^{n-2} 2^{n+1-k} a^k_{1,2} = 2^{n+1}(n-3)\,.
\end{equation}
Similarly,
\begin{equation}\label{eq:app37}
a^n_{3,4} = \sum_{k=4}^{n-2} 2^{n+1-k} a^k_{2,3} = 2^{n+1}(n-4)(n-5)\,.
\end{equation}
Since sites with distances $m$, $m+1$ do not appear before the
construction step $n=2m$, the sum over $a^k_{2,3}$ starts at $k=4$.
Proceeding in this manner, for general $m \ge 1$ and $n \ge 2m$,
\begin{align}
a^n_{m,m+1} &= \sum_{k=2(m-1)}^{n-2} 2^{n+1-k} a^k_{m-1,m}\nonumber\\
&= \frac{2^{m-1} 2^n (n-m-1)!}{(m-1)!(n-2m)!}\,.\label{eq:app38}
\end{align}
The value of $a^n_{0,1}$ is $1$ for $n=0$ and $0$ for $n > 0$.
Analogously, for general $m \ge 2$ and $n \ge 2m-1$,
\begin{equation}
a^n_{m,m} = \frac{2^{m-1} 2^n (n-m-1)!}{(m-2)!(n-2m+1)!}\,.
\label{eq:app39}
\end{equation}
The value of $a^n_{1,1}$ is $2$ for $n=1$ and $0$ for $n > 1$.

Thus we obtain the quantities in Eq.~\eqref{eq:app32},
{\footnotesize
\begin{align}
  N^\text{near}_n &= \sum_{n^\prime=1}^n \sum_{k=1}^{\lfloor n^\prime/2 \rfloor} a^k_{m,m+1}
= \begin{cases} 1+\frac{2}{9}(2^n+4^n-2) \\ 1+\frac{2}{9}(2^n-2)(2^n+1)\end{cases}\nonumber\\
  N^\text{mid}_n &= \sum_{n^\prime=1}^{n} \sum_{k=1}^{\lfloor (n^\prime+1)/2 \rfloor} a^k_{m,m} = \begin{cases}
    \frac{2}{9}(2^n-1)^2 \\
    \frac{2}{9}(2^n+1)^2 \end{cases}\nonumber\\
  d^\text{near}_n &= \sum_{n^\prime=1}^n \sum_{k=1}^{\lfloor n^\prime/2 \rfloor} m a^k_{m,m+1}
= \begin{cases} \frac{2}{81}(2^n+2)(2^n+3\cdot 2^n n -1)\\ \frac{2}{81}(2^n-2)(2^n+3\cdot 2^n n +1)\end{cases}\nonumber\\
  d^\text{mid}_n &= \sum_{n^\prime=1}^n \sum_{k=1}^{\lfloor (n^\prime+1)/2 \rfloor} m a^k_{m,m}\nonumber\\
& = \begin{cases} \frac{2}{81}\left[-2^{n+3}-12\cdot 2^nn+(3n+7)4^n+1\right]\\
    \frac{2}{81}\left[2^{n+3}+7\cdot 4^n +3n(2^{2+n}+4^n)+1\right]\end{cases}\nonumber\\
\label{eq:app40}
\end{align}}where $\lfloor x \rfloor$ denotes the largest integer $\le x$ and the different results for $n$ even and odd are given
consecutively, and {\footnotesize
\begin{align}
  d^\text{far}_n &= \sum_{n^\prime=1}^n \sum_{k=1}^{\lfloor n^\prime/2 \rfloor} (m+1) a^k_{m,m+1} = d_n^\text{near}+ N_n^\text{near}\,.
\label{eq:app40a}
\end{align}}

Substituting the results of Eq.~\eqref{eq:app40} into
Eqs.~\eqref{eq:app33}-\eqref{eq:app35}, for $p=1$,
\begin{align}
  \Delta_n &= \frac{1}{27} \left[-23-8(-2)^n+8\cdot 4^n+(104+48n)
    16^n\right].
\label{eq:app41}
\end{align}

Substituting Eqs.~\eqref{eq:app29} and \eqref{eq:app41} for
$\Delta_m$ into Eq.~\eqref{eq:app8}, and using $S_1 = 8,7$ for
$p=0,1$,
\begin{equation}\label{eq:app9}
  S_n = \begin{cases} \frac{2^n}{189}\left[98+27\cdot 2^n+(42+21n)4^n+22\cdot 16^n\right]\\
    \frac{1}{81}\left[ 23+4(-2)^n +(44+6n)4^n +(10+12n)16^n\right]
  \end{cases}
\end{equation}
Eq.~\eqref{eq:app9} in Eq.~\eqref{eq:app4} yields the analytical
expressions for $\ell_n$ in Eqs.~\eqref{eq:6} and \eqref{eq:7}.

\end{document}